\documentclass[
  reprint,
  preprintnumbers,
  amsmath,amssymb,aps,
  nolong
]{revtex4-2}

\usepackage{graphicx}
\usepackage{dcolumn}
\usepackage{bm}
\usepackage{aas_macros}
\usepackage{hyperref}

\usepackage{graphicx}
\usepackage{xcolor}
\usepackage{multirow}
\usepackage{rotating}
\usepackage[normalem]{ulem}
\usepackage{tikz}

\begin{document}

\preprint{ET-0056A-25}

\title{Detectability of gravitational atoms in black hole binaries with the Einstein Telescope}

\author{Riccardo~Della~Monica}
\email{rdellamonica@usal.es}
\affiliation{Departamento de Física Fundamental, Universidad de Salamanca, Plaza de la Merced, s/n, E-37008 Salamanca, Spain}

\author{Richard~Brito}
\email{richard.brito@tecnico.ulisboa.pt}
\affiliation{CENTRA, Departamento de Física, Instituto Superior Técnico – IST\\
Universidade de Lisboa – UL, Avenida Rovisco Pais 1, 1049-001 Lisboa, Portugal}

\date{\today}

\begin{abstract}
  Rotating black holes can amplify ultralight bosonic fields through superradiance, forming macroscopic clouds known as gravitational atoms. When the cloud forms around one of the components of a binary system, it can undergo a series of distinctive interactions, comprising both secular effects, such as dynamical friction or accretion, and resonant behaviour. These processes are expected to leave a distinctive signature on the gravitational waveform emitted by the binary, whose detectability we investigate in this paper. To do so, we implement a numerical code that integrates these effects,
  computed within a Newtonian approximation,
  for small-to-intermediate mass-ratio binaries on circular equatorial orbits.
  Realistic waveforms incorporating these environmental influences are generated and analyzed using the Fisher matrix formalism to evaluate the detectability of bosonic clouds with current and next-generation
  ground-based
  gravitational wave observatories. Our results demonstrate the potential for gravitational wave astronomy to probe the existence and properties of ultralight bosons.
\end{abstract}

\maketitle

\section{Introduction}
\label{sec:intro}
The first direct detection of gravitational waves \cite{Abbott2016a} in the last decade marked the birth gravitational-wave astronomy, revolutionizing our understanding of compact objects such as black holes \cite{Abbott2016b, Baibhav2021} and offering unparalleled opportunities to explore fundamental physics \cite{Barack2019, Sathyaprakash2019, Bertone2020,Arun2022}. While currently-available observations have significantly advanced our knowledge, offering insights on the population of binary black hole systems \cite{Abbott2023, Edelman2023} and opening new avenues to perform tests of gravity in the strong field regime \cite{Abbott2019, Abbott2021,Abbott2021_2}, the next generation of ground- and space-based detectors, such as the Einstein Telescope (ET) \cite{Maggiore2020, ET2025}, Cosmic Explorer \cite{Evans2021}, and LISA \cite{Colpi2024}, will vastly enhance our ability to study compact binaries and significantly improve current tests of gravity \cite{Perkins2021}.

Due to an overall improvement in sensitivity and a wider accessible frequency band, such next-generation detectors will allow unprecedented tests of beyond General Relativity theories \cite{Huwyler2012, Belgacem2019,Perkins2021, Pan2021,Arun2022,Branchesi:2023mws} and probe the environment surrounding astrophysical black holes \cite{Cardoso2020, Cardoso2022, Cole2023, Zwick2023,CanevaSantoro2024}, like dark matter overdensities \cite{Eda2013, Kavanagh2020, Hannuksela2020,Duque:2023seg} and accretion discs \cite{Yunes2011, Derdzinski2021,Duque:2024mfw}.

A particularly intriguing type of environment is comprised of ultralight bosons in the mass range $10^{-21} - 10^{-11}$ eV \cite{Hui2021}, usually regarded as compelling and well-motivated dark matter candidates \cite{Hui2017}. While such bosons have not been detected so far, they can emerge in theories beyond the Standard Model and may offer solutions to unresolved issues in particle physics and astrophysics \cite{Abbott1983, Schive2014, Marsh2015, DellaMonica2023a, DellaMonica2023b, DellaMonica2024}. Examples include the QCD-axion \cite{Peccei1977} and axion-like fields from string compactifications \cite{Svrcek2006, Arvanitaki2010}. If a massive bosonic field exists in the vicinity of a rotating black hole, they can trigger an instability~\cite{Damour:1976kh,Detweiler1980,Cardoso:2005vk,Dolan:2007mj} and extract energy and angular momentum from the rotating black hole through the classical superradiance mechanism \cite{Zeldovich1971,Press1972, Starobinski1973, Brito2015}. In this process, up to $\sim10\%$ of the black hole's energy can be transferred to the bosonic field \cite{Herdeiro2022}. As a result, the black hole spins down \cite{Brito2015b, East:2017ovw} and becomes surrounded by a long-lived cloud of ultralight bosons that is often referred to as a ``gravitational atom'', due to the similarity in the mathematical description of these systems with that of an hydrogen atom.

Several mechanisms have been proposed to probe the existence and formation of boson clouds. One possibility is related to the superradiantly-induced spindown of rotating black holes, which could result in the absence of highly spinning black holes within specific mass ranges, depending on the boson mass~\cite{Arvanitaki:2010sy}. By looking for gaps in the mass-spin diagram of astrophysical black holes one could infer or place constraints on the existence of ultralight bosons \cite{Arvanitaki2014, Brito2015b}. From observations of stellar-mass black holes, this methodology currently sets an excluded region for the boson mass in the range $\sim [10^{-13},\,10^{-12}]$ eV~\cite{Arvanitaki2014, Cardoso2018,Ng2020,Cheng:2022jsw,Hoof:2024quk}, when neglecting non-gravitational interactions.
Another interesting prospect is the direct detection of long-lived, nearly monochromatic gravitational waves emitted by boson clouds through pair annihilation within the cloud \cite{Yoshino2014,Arvanitaki2014,Brito2017b,Isi:2018pzk}. Such radiation could manifest as either a continuous signal from individual sources or as a stochastic gravitational background \cite{Arvanitaki2014, Brito2017, Brito2017b}.
Current all-sky searches of a continuous gravitational-wave signal from boson clouds using LIGO, VIRGO and KAGRA \cite{Palomba2019,Zhu2020,Abbott2021c} as well as stochastic background searches~\cite{Tsukada:2018mbp,Tsukada:2020lgt,Yuan2022}, and the subsequent lack of detection, also disfavors ultralight bosons that only interact gravitationally in the mass range $\sim [10^{-13},\,10^{-12}]$ eV.

Alternatively, a potential observational avenue to search for signatures from ultralight bosons, that we will explore in this work, is given by clouds that are formed around black holes in coalescing binaries: interactions between the bosonic cloud and the binary companion give rise to a rich phenomenology \cite{Baumann2019, Zhang2019, Zhang2020, Baumann2020, Baumann2022,Takahashi2023, Tomaselli2023,Tomaselli2024,Tomaselli2024b,Boskovic2024,Kyriazis:2025fis}. As the companion spirals inward, it scans increasingly high orbital frequencies, exciting distinct interaction channels in the system. For example, at specific orbital frequencies, matching the difference in phase velocities of different modes of the cloud, the gravitational perturbations produced by the companion is resonantly amplified \cite{Baumann2019,Baumann2020}. This resonance forces the cloud to transition — partially or entirely — between states. These resonant transitions of the cloud have a strong backreaction on the binary orbit, which has to compensate for the change in energy and angular momentum of the cloud, causing the inspiral to either stall or accelerate \cite{Takahashi2023, Tomaselli2024}. While this process leaves a potentially detectable dephasing imprint in the gravitational wave signal of the binary, it can also lead to an early depletion of the cloud, before the systems enter the detectors bands \cite{Baumann2019,Berti2019, Baumann2020,Tomaselli2024}. Nonetheless, it has been shown that when astrophysical processes like the common envelope phase \cite{Guo2024} are taken into account, or for a large range of binary parameters that make the cloud unaffected by these early resonances \cite{Tomaselli2024}, the cloud can survive until later stages of the inspiral.

Further down the merger history, when the binary system nears merger and its separation approaches the size of the cloud, the gravitational perturbation from the companion induces a different type of transition to unbound states of the cloud, analogous to ionization of atoms \cite{Baumann2022, Tomaselli2023}. This process is powered by the binary's energy, which experiences dynamical friction \cite{Vicente2022} leading to an energy loss that can largely exceed the one resulting from gravitational wave emission. As a result, in this phase the dynamics of the inspiral is dominated by the interaction with the cloud rather than by gravitational radiation reaction.
Moreover, ionization is also characterized by discontinuities when the orbital frequency crosses certain thresholds. These \emph{sharp} features \cite{Baumann2022b} leave potentially observable imprints on the gravitational waveform, providing potential direct insight into the microscopic structure of the cloud.

Finally, due to the high typical densities of such clouds the accretion of the boson cloud onto the companion also plays a significant role \cite{Vicente2022, Traykova2023, Baumann2022}. In some cases, the increase in the companion mass can be associated with a substantial acceleration of the merger relative to the evolution in vacuum, representing another possible observational signature.

This work builds upon past studies in which all such interaction effects have been thoroughly investigated in the non-relativistic regime \cite{Baumann2020, Baumann2022,Takahashi2023, Tomaselli2023,Tomaselli2024}, providing a solid theoretical background for our study. Moreover, recent work on boson clouds in extreme mass-ratio binary systems \cite{Duque:2023seg,Brito2023,Dyson2025}, making use of black hole perturbation theory, provided an independent analysis of the binary-cloud interactions, this time in a fully relativistic regime. These results seem to be in agreement with those from the non-relativistc counterpart, at least at a qualitative level, further consolidating the picture that we have described above. Here, we address the problem of the potential future detectability of boson clouds through their signature on the gravitational waveforms emitted during inspiral, using next-generation gravitational wave ground-based observatories, such as the ET.
To do so, we implement the above mentioned theoretical results in a numerical framework that takes into consideration the full interplay between the different interaction effects experienced by the cloud-binary system, in a comprehensive and systematic manner. For simplicity, we restrict ourselves to the case of circular equatorial orbits,
leaving the study of more generic orbits to future work.
Our numerical code allows us to compute the full evolution of such systems in the presence of a scalar boson cloud and to estimate the amount of dephasing in the emitted gravitational wave signal with respect to a vacuum environment. Using a phenomenological waveform model for the inspiral, merger and ringdown of quasi-circular non-precessing black hole binaries and our dephasing estimate, we compute realistic waveforms including the environmental effects from the boson cloud. These waveforms are then used to assess the detectability of the cloud with current and future ground-based detectors and the ability to estimate the cloud parameters. This is done by performing both a mismatch analysis and a Fisher Matrix analysis. Considering the full interplay between cloud-binary interaction effects, our work complements past detectability studies of binary-cloud systems in the literature \cite{Cole2023,Khalvati2024} which only focused on space-based detectors such as LISA, representing a step forward not only in terms of a broader region of the parameters space explored, but also on the modeling side.

The rest of this paper is organized as follows: in \S\ref{sec:model} we present a comprehensive review of the literature on the interaction effects within boson cloud-binary systems, we give an overview of the theoretical model that we implemented in our numerical framework and how we used it to determine realistic waveforms for these systems. In \S\ref{sec:fisher_matrix} we briefly review the Fisher Matrix formalism and show the results obtained with our model. Finally, \S\ref{sec:discussion} is devoted to discussion and conclusions. For most of the this work, we use geometric units $G = c = \hbar =  1$, unless otherwise stated.

\section{Model}
\label{sec:model}

In this section we explore the details of the model we used for the generation of gravitational waveforms for the inspiral of compact binaries in the presence of an ultralight boson cloud. First, in \S\ref{sec:ultralight_bosons} we give an overview of how, via the supperadiance instability, rotating black holes can trigger the formation of macroscopic boson clouds in their surroundings. Then, in \S\ref{sec:binary_interactions} we will show how the tidal perturbation, induced by the presence of a binary companion, can transform the cloud's configuration implying a transfer of energy and angular momentum which, in turn, produces a backreaction on the orbit. In \S\ref{sec:orbit_evolution} we describe our numerical methodology to take into account such backreaction in the orbital evolution of the binary system during inspiral and show an example evolution for a benchmark model. Finally, in \S\ref{sec:waveforms} we describe how, from the integrated orbits, we can reconstruct realistic gravitational waveforms for the signal emitted by such systems.

\subsection{Ultralight boson clouds}
\label{sec:ultralight_bosons}

\begin{figure*}[!t]
  \includegraphics[width=\textwidth]{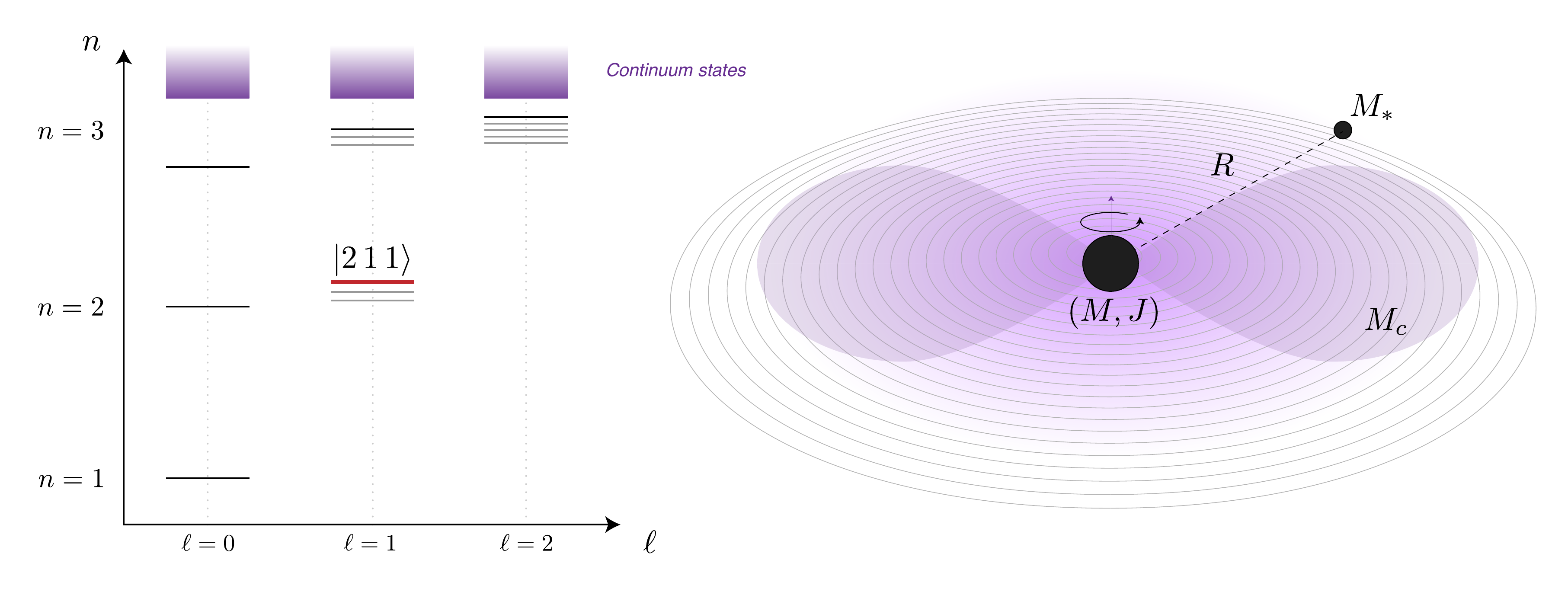}
  \caption{Spectrum of the bound states of a gravitational atom for the first few values of $n$. The highlighted $|211\rangle$ state is the fastest growing mode and its corresponding mass distribution for the real case
    on the equatorial plane is shown in the illustration on the right panel, where we also depict the geometrical configuration of the binary system at a separation $R$. The central spinning black hole, with mass $M$
    and angular momentum $J$,
  is surrounded by the boson cloud of mass $M_c$. The tidal perturbation induced by the presence of the binary companion, with mass $M_*$, leads to a momentum and energy transfer between the cloud and the binary, whose signature on the gravitational wave signal we seek to study.}
  \label{fig:boson-cloud-illustration}
\end{figure*}

Bosonic fields can extract energy and angular momentum from rotating black holes through a process known as \textit{superradiance} \cite{Brito2015}, similar to the Penrose process. This phenomenon takes place when the boson's angular frequency, $\omega_B$, is smaller than the angular velocity of the event horizon, $\Omega_H$, specifically when
\begin{equation}
  \omega_B < m\Omega_H,
  \label{eq:superradiant-regime}
\end{equation}
being $m$ the azimuthal quantum number in the black hole frame. If the bosons are massive, the superradiantly amplified waves can become confined around the black hole, leading to an exponential increase in their occupancy number and to the formation of macroscopic boson clouds. For maximal efficiency in superradiant amplification, the gravitational radius $r_g$ of the black hole and the Compton wavelength $\lambda_c$ of the field need to be comparable,
\begin{equation}
  \alpha \equiv \frac{r_{g}}{\lambda_{c}} = \mu M \lesssim 1
  \label{eq:fine-structure}
\end{equation}
where $M$ and $\mu$ are the masses of the central black hole and of the bosonic particles, respectively, and the ratio $\alpha$ is a dimensionless constant commonly known as the ``gravitational fine structure constant''.

The dynamics of a massive scalar field $\Phi$ with mass $\mu$ in a curved spacetime is given by the Klein-Gordon equation
\begin{equation}
  \Bigl(g^{\alpha\beta}\nabla_\alpha\nabla_\beta-\mu^2\Bigr)\,\Phi(t, \vec{r}) = 0\,,
  \label{eq:Klein-Gordon}
\end{equation}
where $g_{\alpha\beta}$ is the spacetime metric  (in our case the Kerr metric) and $\nabla_{\alpha}$ denotes the covariant derivative. In the non-relativistic limit (equivalent to $\alpha \ll 1$), one can find quasi-bound solutions resembling those of the hydrogen atom in quantum mechanics. Assuming that one can factor out the rapid variability of the scalar field on timescales $\mu^{-1}$, one can use the ansatz
\begin{equation}
  \Phi(t, \vec{r})=\frac{1}{\sqrt{2 \mu}}\bigl[\psi(t, \vec{r}) e^{-i \mu t}+\psi^*(t, \vec{r}) e^{+i \mu t}\bigr]\,.
  \label{eq:ansatz-scalar-field}
\end{equation}
In this case, the Klein-Gordon equation for the slowly varying component of the scalar field, $\psi(t,\vec{r})$, can be recast into a Schrödinger equation with a Coulomb-like potential
\begin{equation}
  i \frac{\partial}{\partial t} \psi(t, \vec{r})=\biggl[-\frac{1}{2 \mu} \nabla^{2}-\frac{\alpha}{r}\biggr] \psi(t, \vec{r})\,.
  \label{eq:Schrodinger}
\end{equation}

Assuming separation of variables, the solutions to Eq.~\eqref{eq:Schrodinger} can be expressed as
\begin{align}
  |n\ell m\rangle \equiv \psi_{n \ell m}(t, \vec{r})=R_{n \ell}(r) Y_{\ell m}(\theta, \phi) e^{-i\left(\omega_{n \ell m}-\mu\right) t}\,.\label{eq:eigenstates}\\
  \nonumber
\end{align}
Here, $Y_{\ell m}$ are spherical harmonics, and $R_{n\ell}$ are hydrogen-like radial functions, labeled in terms of $n$, $\ell$, and $m$, which correspond to the principal, angular momentum, and azimuthal quantum numbers, respectively, satisfying the usual relations found in the hydrogen atom: $n > \ell$, $\ell \geq 0$ and $\ell \geq |m|$ \cite{Baumann2019b}. The main distinction from the hydrogen atom is the boundary condition at the black hole horizon, leading to quasi-bound states with complex eigenfrequencies
\begin{equation}
  \omega_{n \ell m} = \omega_{n\ell m,\,R}+i\,\omega_{n\ell m,\,I}\,.
  \label{eq:complex-eigenfrequency}
\end{equation}
When the superradiant condition, Eq.~\eqref{eq:superradiant-regime}, is satisfied the imaginary part of the eigenfrequency is positive \cite{Detweiler1980,Baumann2019b}, leading to an exponential growth of the amplitude of the unstable mode. During the superradiant phase the central black hole transfers mass and angular momentum to the boson cloud, thus reducing its angular velocity. This eventually leads to a saturation of the condition in Eq.~\eqref{eq:superradiant-regime}, which marks the end of the instability~\cite{Brito2015b,East:2017ovw}. The endpoint of this process is a Bose-Einstein condensate around the black hole populating one (or more) of the modes in Eq.~\eqref{eq:eigenstates}, a system referred to as a ``gravitational atom''. The fastest-growing state is the $|211\rangle$, and it is estimated that the cloud mass, $M_c$, can reach up to $10\%$ of the central black hole mass, very quickly on astrophysical timescales. The exact value of $M_c$ depends on the initial mass and spin of the primary black hole, and on the subsequent evolution of the boson cloud, which, in general, is not known a priori. Hence, in this work, we take the agnostic approach of leaving $M_c$ as a free parameter.

For $\alpha \ll 1$, one can write the real part of the eigenfrequencies as \cite{Baumann2019b}
\begin{widetext}
  \begin{equation}
    \epsilon_{n\ell m}\equiv \omega_{n\ell m,\,R} =\mu\left(1-\frac{\alpha^{2}}{2 n^{2}}-\frac{\alpha^{4}}{8 n^{4}}-\frac{(3 n-2 \ell-1) \alpha^{4}}{n^{4}(\ell+1 / 2)}+\frac{2 \tilde{a} m \alpha^{5}}{n^{3} \ell(\ell+1 / 2)(\ell+1)}+\mathcal{O}\left(\alpha^{6}\right)\right)\,,
    \label{eq:eigenenergy}
  \end{equation}
\end{widetext}
which form a discrete spectrum of energy levels (see Fig.~\ref{fig:boson-cloud-illustration}).
Here $\tilde{a}\equiv J/M^2$ is the dimensionless black-hole spin parameter, being $J$ its angular momentum.
The energy difference between distinct modes presents the same hierarchy found in the hydrogen atom, with the greatest jump corresponding to a Bohr splitting ($\Delta n \neq 0$), a much smaller jump at a fine splitting level ($\Delta n = 0$, $\Delta \ell \neq 0$ appearing at $\mathcal{O}(\alpha^4)$) and an hyperfine splitting ($\Delta n = 0$, $\Delta \ell = 0$ and $\Delta m \neq 0$) corresponding to the smallest energy jump, appearing at order $\mathcal{O}(\alpha^5)$.

In the case of a cloud made up of a real scalar field, with time dependent and non-axisymmetric stress-energy tensors, the quasi-bound states eventually dissipate over time due to gravitational wave emission sourced by pair annihilation within the cloud \cite{Yoshino2014, Arvanitaki2014,Brito2015b} (in the case of a complex field the cloud does not dissipate through gravitational wave emission~\cite{Herdeiro:2014goa}). The frequency of the continuous gravitational waves emitted  by a given mode $|n\ell m\rangle$ is given by $f_\textrm{GW} = \omega_{n\ell m, R}/\pi$. In particular, in the small$-\alpha$ limit, the energy flux emitted by the $|2\,1\,1\rangle$ mode is approximately given by \cite{Brito2015b}
\begin{equation}
  \dot{E}_\textrm{GW, cloud} \approx \frac{484+9\pi^2}{23040}\left(\frac{M_c}{M}\right)^2\alpha^{14}\,,
  \label{eq:continuous-gw-emission_dotE}
\end{equation}
The depletion timescale of the cloud due the gravitational emission is always much larger than the superradiance instability timescale, and grows with the quantum number $m$, so that higher-order modes dissipate slower \cite{Yoshino2014}. Hence, while the dissipation does not inhibit the formation of the cloud, the clouds may still deplete on cosmological/astrophysical timescales (especially for large $\alpha$, due to the strong dependence of Eq.~\eqref{eq:continuous-gw-emission_dotE} from this parameter).
More concretely, for an isolated black hole, the cloud decays through gravitational wave emission as
\begin{equation}
  M_c(t) \approx \frac{M_{c,0}}{1+t/\tau_{\rm gw}}\,,
  \label{eq:Mcgw}
\end{equation}
where $M_{c,0}\equiv M_c(t=0)$ is the mass of the cloud at the time of formation and the gravitational-wave timescale for dissipation is
\begin{equation}
  \tau_{\rm gw} \approx 2.5\times 10^6\left(\frac{0.01}{M_{c,0}/M}\right)\left(\frac{M}{40 M_{\odot}}\right)\left(\frac{0.1}{\alpha}\right)^{14}\, {\rm yr}\,.
  \label{eq:continuous-gw-emission}
\end{equation}
This approximation slightly underestimates the correct dissipation timescale computed numerically~\cite{Yoshino2014, Siemonsen:2022yyf}, especially at large $\alpha$, however it allows us to give an estimate of the maximum value of $M_c(t)$ that one expects given the formation age. For example, for a $M=40M_{\odot}$ black hole and assuming $M_{c,0}\approx 10^{-2}M $ and $\alpha=0.1$ one would have $M_c(t=20{\rm Myrs})\approx 10^{-3}M$.

In addition to gravitational-wave dissipation, when considering a realistic evolution of cloud, we should also take into account the fact that on long timescales the black hole will keep spinning down due to the growth of higher $m$ modes. Due to the hierarchy of the instability timescales for different $m$ modes, the growth of each mode typically proceeds in a step-wise fashion, in which the cloud is typically in a nearly pure state at each step~\cite{Ficarra:2018rfu,Yuan:2021ebu,Guo:2022mpr}. For example, assume that a black hole is born with a sufficiently large spin, meaning $\tilde{a}\gtrsim 4m\alpha/(m^2+4\alpha^2)$ for a given mode $m$, such that the first mode to grow is the $|211\rangle$. For $\alpha=0.1$ and $M=40M_\odot$ the e-folding instability timescale $\tau_{\rm inst}^{211}$ for the $|211\rangle$ mode to grow would be $\tau_{\rm inst}^{211}\lesssim 50$ yr, as long as $\tilde{a}\gtrsim 0.39$, the exact timescale being dependent on the initial black hole spin (note that this timescale depends on $\alpha$ roughly as $\propto\alpha^{-4l-5}$~\cite{Detweiler1980,Baumann2019b,Dolan2018}). During the process the black hole spins down until saturating at $\tilde{a}\sim 4\alpha/(1+4\alpha^2)$, which typically occurs after about 200 e-folding times~\cite{Ficarra:2018rfu,Zhu2020,Guo:2022mpr}. At that point, the black hole parameters remain roughly constant for some time, with the cloud dissipating through gravitational-wave emission, until the next most unstable mode starts growing. In our case this would be the $|322\rangle$ mode, which at that point would grow on a timescale
\begin{equation}
  \tau^{322}_{\rm inst}\approx 10^7\left(\frac{M}{40 M_{\odot}}\right)\left(\frac{0.1}{\alpha}\right)^{14}\, {\rm yr}\,,\label{eq:inst_322}
\end{equation}
where we used the analytical approximation in~\cite{Baumann2019b} at leading order in $\alpha$ and fixed the black hole spin to be $\tilde{a}\sim 4\alpha/(1+4\alpha^2)$. Comparing Eq.~\eqref{eq:inst_322} with Eq.~\eqref{eq:continuous-gw-emission} one sees that, typically, the $|211\rangle$ dissipates on a timescale that is similar to the one needed for the next most unstable mode to grow, meaning that once the $|322\rangle$ starts becoming relevant there is typically a negligible amount of energy in the $|211\rangle$ mode~\cite{Ficarra:2018rfu,Guo:2022mpr}. If there is no gravitational-wave dissipation, as can happen for a complex scalar field~\cite{Herdeiro:2014goa}, the $|211\rangle$ would quickly be absorbed by the black hole once the $|322\rangle$ starts growing \cite{Ficarra:2018rfu, Hui2023}. For concreteness, below we will only consider black holes surrounded by a cloud in a $|211\rangle$ state, leaving the study of higher modes for future work. In this scenario, the timescales in Eqs.~\eqref{eq:continuous-gw-emission} and~\eqref{eq:inst_322} provide a rough timescale for the maximum age that the black hole with the cloud can have at the time at which we observe the binary. Much older black holes would be expected to be surrounded by clouds in higher-$m$ states. While specific estimates would require an in-depth study of different formation channels, we note that in some scenarios binaries can form and merge in timescales of $\mathcal{O}(10\rm{Myr})$~\cite{duBuisson,diCarlo}. When considering realistic values for $\alpha$ and $M_c$ for an observed binary black hole system with a cloud in a $|211\rangle$ state one should take these caveats into consideration. Namely, observing such a cloud with $\alpha \lesssim 0.1$ is much more likely than for larger values of $\alpha$. We will nevertheless extrapolate our results to larger values of $\alpha$ in order to explore the parameter space in detail. This will also allow us to show that, in contrast to LIGO and Virgo, clouds in a $|211\rangle$ state could be detectable in ET for values of $M_c$ as small as $M_c\sim \mathcal{O}(10^{-3})$ and $\alpha\lesssim 0.1$.

Finally, below we also make use of the fact that the spectrum
of Eq.~\eqref{eq:Schrodinger}
also includes unbound states described by \cite{Baumann2019b}
\begin{equation}
  |k;\,\ell m\rangle \equiv \psi_{k; \ell m}(t, \vec{r})=R_{k; \ell}(r) Y_{\ell m}(\theta, \phi) e^{-i\epsilon(k)t}\,,
  \label{eq:unboundstates}
\end{equation}
where $k$ is a positive, real-valued continuous wavenumber. The radial function for these states reads as
\begin{widetext}
  \begin{equation}
    R_{k ; \ell}(r)=\frac{2 k e^{\frac{\pi \mu \alpha}{2 k}}\bigl|\Gamma\bigl(\ell+1+\frac{i \mu \alpha}{k}\bigr)\bigr|}{(2 \ell+1) !}(2 k r)^{\ell} e^{-i k r}\,{ }_{1}F_{1}\biggl(\ell+1+\frac{i \mu \alpha}{k} ; 2 \ell+2 ; 2 i k r\biggr)\,,
    \label{eq:radialcontinuum}
  \end{equation}
\end{widetext}
being $_1F_1$ the Kummer confluent hypergeometric function. The exact eigenfrequencies  for this unbound states are purely real and follow the dispersion relation
\begin{equation}
  \epsilon(k) = \sqrt{\mu^{2}+k^{2}} - \mu \approx \frac{k^{2}}{2\mu}\,,
  \label{eq:dispersionrelation}
\end{equation}
valid in the non-relativistic limit ($k \ll \mu$). Similarly to the hydrogen atom case, these states represent the situation in which the scalar field is not bound to the black hole, and can thus be thought of as scattering states.

\subsection{Tidal perturbation in a binary system}
\label{sec:binary_interactions}

\begin{figure*}[!t]
  \includegraphics[width=\textwidth]{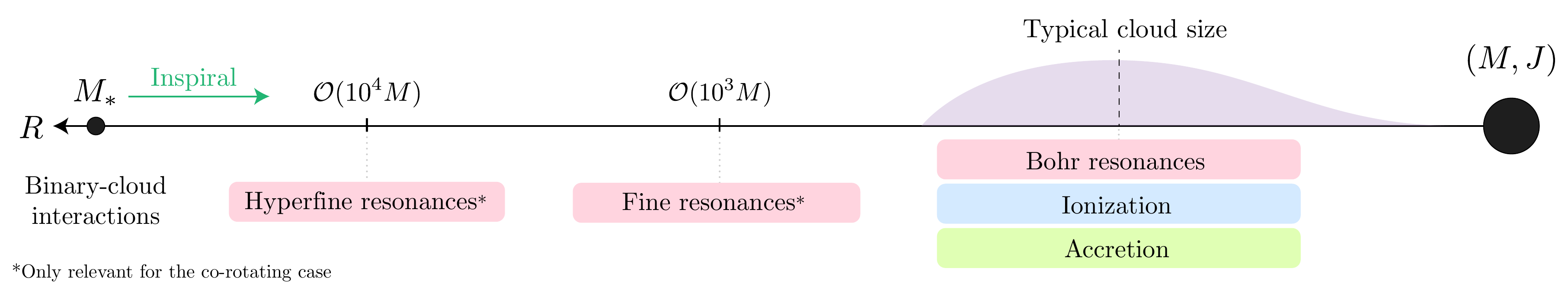}
  \caption{
  Illustration of the chronological (\emph{i.e.} as a function of the decreasing binary separation during the inspiral) sequence of effects related to the tidal interaction between the boson cloud and the binary system. The earliest interactions correspond to hyperfine (excited at $R\sim\mathcal{O}(10^4M)$ for the case $M\mu=0.2$) and fine (at $R\sim\mathcal{O}(10^3M)$ for the same choice of parameters) bound-level mixing, occurring in the co-rotating case. If the cloud survives these resonances (which can happen for some combination of parameters) or if the binary-cloud configuration is counter-rotating, additional interactions are enabled when the binary separation is on the same order as the cloud size: namely the ionization of the cloud, resulting in a dynamical friction effect on the binary orbit, the accretion of the cloud on the secondary object and Bohr resonances. All such effects are studied in Section \ref{sec:binary_interactions}.}
  \label{fig:interactions}
\end{figure*}

In this work, we consider the case in which the boson cloud forms around the heaviest component of an asymmetric black hole binary system. In particular, denoting as $M$ the mass of this object and as $M_{*}$ the mass of its companion, we consider mass ratios $q \equiv M_*/M \ll1$. Using a reference frame centered on the heaviest black hole, we identify the position of the companion by $\vec{R}_{*} =(R,\,\theta_*,\,\varphi_{*})$, being $R$ the binary's separation and $\theta_*$ the polar angle from the direction of the black hole's spin.
While non-circular and inclined orbits have been analyzed in literature \cite{Tomaselli2023, Tomaselli2024, Boskovic2024,Tomaselli2024b}, we restrict ourselves for simplicity to the case of circular equatorial orbits,
identified by $\theta_* = \pi/2$ with zero eccentricity. While these assumptions limit the generality of our results, they allow us to describe the dynamics of the system and the different interaction effects in a much simpler way. Moreover, since we are interested in studying the emission of gravitational waves in the band of ground-based detectors (\emph{i.e.} in the last phase of the inspiral) we can safely assume that by that time the circularizing effect of gravitational wave emission \cite{Peters1964} has
substantially
reduced any non-zero eccentricity in the early phases of the inspiral. Additionally, the ionization of the gravitational atom, which we will introduce in more detail in the next sections, further suppresses the orbital eccentricity \cite{Tomaselli2023}, further justifying our assumptions.
We will discuss the implications and limitations of these assumptions in \S\ref{sec:discussion}.

To leading-order in a post-Newtonian expansion, the binary system emits gravitational waves through the standard general relativistic quadrupolar emission, bringing the orbital frequency $\Omega(t)$ to slowly increase (chirp) over time. In the absence of a surrounding environment, this is the only contribution that drives the inspiral and merger of the binary system, whose instantaneous rate of energy loss is given by \cite{Peters1964}
\begin{align}
  \label{eq:p_gw}
  P_{\rm GW} &= \frac{32}{5}\frac{q^{2}M^{5}(1+q)}{R_*^{5}}\,.
\end{align}
If a boson cloud is present around the central black hole, the massive companion interacts with it gravitationally, perturbating its state, and producing a backreaction on the orbit that modifies the inspiral evolution. To describe how the cloud responds to the presence of the massive perturber, it is useful to write down its Newtonian gravitational potential, $V_*$, as a multipole expansion,
\begin{widetext}
  \begin{equation}
    V_*(t,\vec r)=-\sum_{\ell_*=0}^\infty\sum_{m_*=-\ell_*}^{\ell_*}\frac{4\pi q\alpha}{2\ell_*+1}Y_{\ell_*m_*}(\theta_*,\varphi_*)Y_{\ell_*m_*}^*(\theta,\phi)\,F(r)\,,
    \label{eq:V_star}
  \end{equation}
  where
  \begin{equation}
    F(r)=
    \begin{cases}
      \dfrac{r^{\ell_*}}{R_*^{\ell_*+1}}\Theta(R_*-r)+\dfrac{R_*^{\ell_*}}{r^{\ell_*+1}}\Theta(r-R_*)&\text{for }\ell_*\ne1\,,\\[12pt]
      \biggl(\dfrac{R_*}{r^2}-\dfrac{r}{R_*^2}\biggr)\Theta(r-R_*)&\text{for }\ell_*=1\,,
    \end{cases}
  \end{equation}
\end{widetext}
and $\Theta$ corresponds to the Heaviside function. The potential in Eq.~\eqref{eq:V_star} induces a perturbation that leads to a mixing between the cloud's states. In particular, mixing between bound states, happening at specific orbital frequencies, can bring a fraction of the cloud from its original mode to others in the spectrum,
through a \textit{resonant} transition. Mixing between bound and unbound states, on the other hand, can partially \emph{ionize} the cloud, carrying energy and angular momentum from the system to infinity. Moreover, since the secondary object is a black hole, absorption of a fraction of the cloud, which we refer to as \emph{accretion}, plays an important role in the binary's evolution. The different interactions between the binary and the boson cloud turn on at different orbital separations during the inspiral, as schematically illustrated in Fig.~\ref{fig:interactions}.

Overall, neglecting the spin of the binary companion, the evolution of the system is described by five dynamical variables, namely
\begin{equation}
  (R,\,M_c,\,M_*, M,\,J),
  \label{eq:dynamical_system}
\end{equation}
regulated by the system of differential equations
\begin{align}
  \dot{E}_{\rm orb} &= -P_{\rm GW} - P_{\rm ion} - P_{\rm acc} - P_{\rm res}\,, \label{eq:E_orb_evolution}\\
  \dot{M}_c &= -\dot{M}_{{\rm *, acc}} + \dot{M}_{c,{\rm ion}} + \dot{M}_{c,{\rm res}}\,,\label{eq:M_c_evolution}\\
  \dot{M}_* &= \dot{M}_{\rm *, acc}\,,\label{eq:M_star_evolution}\\
  \dot{M} &= \dot{M}_{\rm res}\,,\label{eq:M_evolution}\\
  \dot{J} &= \dot{J}_{\rm res}\,,\label{eq:J_evolution}
\end{align}
being $\dot{E}_{\rm orb}$ the rate of change of orbital energy, related to the change in orbital separation through
\begin{equation}
  \dot{E}_{\rm orb} = \frac{qM^2}{2R^2}\frac{dR}{dt}\,,
\end{equation}
where we have discarded terms related to the variation of $q$ and $M$ which are sub-dominant in the binary evolution. The different terms in Eqs.~\eqref{eq:E_orb_evolution}-\eqref{eq:J_evolution} correspond to each of the possible interaction effects between cloud and binary system. We will now briefly review all such effects. Throughout this section, we will always work in the limit $\alpha\ll1$, which will allow us not to consider relativistic corrections to our model. For this reason, the extrapolations that we will make for values of $\alpha \gtrsim 0.2$ should be interpreted with care.

\paragraph*{Resonances}
As first demonstrated in \cite{Baumann2019, Baumann2020}, when the companion perturbs the cloud at a gradually increasing frequency, transitions between modes occur, similar to bound-bound transitions for a hydrogen atom in quantum mechanics. Since in the case of equatorial quasi-circular orbits the interaction between the binary and the cloud, encoded in the gravitational potential perturbation in Eq.~\eqref{eq:V_star}, oscillates quasi-periodically at a well-defined orbital frequency $\Omega$, a transition channel between any two bound states $|n_a\ell_am_a\rangle$ and $|n_b\ell_bm_b\rangle$ of the cloud is resonantly enhanced only when $\Omega$ is close to the difference between the phase velocities of the two modes, corresponding to the resonance frequency
\begin{equation}
  \Omega_\textrm{res} = \pm\frac{\Delta \epsilon}{\Delta m} > 0,
  \label{eq:omega_res}
\end{equation}
with $\Delta\epsilon \equiv \epsilon_{n_b\ell_bm_b}-\epsilon_{n_a\ell_am_a}$ and $\Delta m \equiv m_b-m_a$,
and the $+$ ($-$) sign applying to co- (counter-) rotating orbital configurations.
Since both $\Delta\epsilon$ and $\Delta m$ can be either positive or negative
and $\Omega_\textrm{res} > 0$,
the configuration of the orbit dictates which resonances are excited: orbits that co-rotate with the cloud can only excite resonances having $\Delta\epsilon$ and $\Delta m$ with same sign, counter-rotating orbits, on the other hand, are allowed when their signs are opposite. The strength of such resonances, which ultimately regulates the fraction of mass of the boson cloud transitioning from one state to the other, is encoded in the level-mixing, defined as the matrix element
\begin{equation}
  \langle n_a\ell_am_a|V_*|n_b\ell_bm_b\rangle = -\sum_{\ell_*m_*}\frac{4\pi\alpha q}{2\ell_*+1}Y_{\ell_*m_*}(\theta_*,\phi_*)I_r I_\Omega
  \label{eq:bound-bound-mixing}
\end{equation}
where $I_r$ and $I_\Omega$ are integrals over radial and angular variables as defined in \cite{Baumann2019,Baumann2022, Tomaselli2023}, that are non-zero only when the selection rules
\begin{align}
  &m_b-m_a = m_* \label{eq:selection_1}\\
  &\ell_a+\ell_*+\ell_b = 2p\,\qquad \textrm{with }p\in\mathbb{Z}\label{eq:selection_2}\\
  &|\ell_b-\ell_a| \leq\ell_*\leq\ell_b+\ell_a\label{eq:selection_3}
\end{align}
are satisfied. Due to the quasi-periodicity of the azimuthal coordinate $\phi_*$ of the binary companion, the matrix element in Eq.~\eqref{eq:bound-bound-mixing} is an oscillatory function that can be written as a sum of overtones
\begin{equation}
  \langle n_a\ell_am_a|V_*|n_b\ell_bm_b\rangle = \sum_{g\in\mathbb{Z}}\eta^{(g)}e^{ig\phi_*}.
  \label{eq:eta_g}
\end{equation}
For equatorial quasi-circular orbits that we are considering the only non-zero term in Eq.~\eqref{eq:eta_g} is $g = \pm \Delta m$ (with $+$ for the co-rotating configuration and $-$ for the counter-rotating one) and the mixing term  $\eta^{(g)}$ depends on time only through the orbital separation $R_*$ (or, equivalently, the orbital frequency $\Omega$) which shrinks over time due to the emission of gravitational waves and all the interaction effects between the cloud and the binary.

\begin{figure*}
  \includegraphics[width=0.8\textwidth]{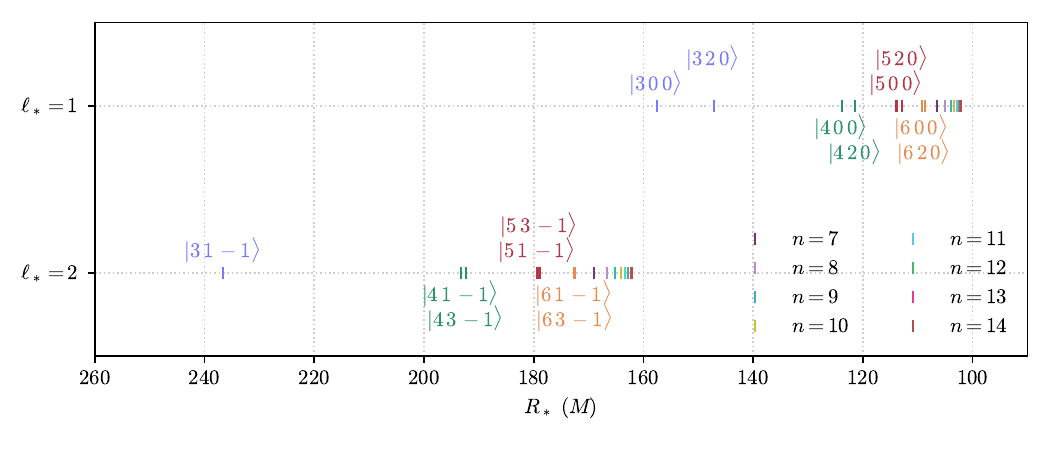}
  \caption{Binary separations at which transitions of the cloud in a $|2\,1\,1\rangle$ state to states with $n_b\geq3$ (shown here up to $n_b=14$) are resonantly excited by the dipole ($\ell_*=1$) and quadrupole ($\ell_*=2$) couplings, for a
    counter-rotating
    binary with $q = 0.1$
    and $\alpha=0.2$.
  The selection rules in Eqs.~\eqref{eq:selection_1}-\eqref{eq:selection_3} make it so that, except for the transition to the $|3\,1\,\textrm{-1}\rangle$, resonances come in doublets (to $|n\,0\,0\rangle$ and $|n\,2\,0\rangle$ states for the dipole and to $|n\,1\,\textrm{-1}\rangle$ and $|n\,3\,\textrm{-1}\rangle$ for the quadrupole) at very close orbital separations between each other. Nevertheless, in each doublet the transition to the second of the two states are strongly suppressed due to the high decay width of the excited states.}
  \label{fig:resonances}
\end{figure*}

The mixing term, $\eta^{(g)}$ not only defines the strength of the resonant transition, but also the amplitude of the frequency bandwidth where the resonance occurs. In all realistic cases, $\Delta\Omega\sim\eta^{(g)}$ is much narrower than the frequency separation between resonances involving different states \cite{Baumann2019, Tomaselli2024}. This allows to treat resonances in the two-state approximation (for a given orbital frequency only transitions between the long-lived cloud state and one excited state are considered) which is described with an analogous formalism to the Landau-Zener mechanism of quantum mechanics. In this case, the time evolution of particle number in the two modes is given by
\begin{equation}
  \frac{d}{dt}
  \begin{pmatrix}
    c_a\\
    c_b
  \end{pmatrix} = -i \mathcal{H}
  \begin{pmatrix}
    c_a\\
    c_b
  \end{pmatrix},
\end{equation}
where the interaction Hamiltonian, in a ``dressed'' frame \cite{Baumann2020} that factors out the rapidly oscillating terms, is given by
\begin{equation}
  \mathcal{H} =
  \begin{pmatrix}
    -(\Delta\epsilon-g\Omega)/2 & \eta^{(g)}\\
    \eta^{(g)} & (\Delta\epsilon-g\Omega)/2
  \end{pmatrix}.
\end{equation}
When the initial state of the system is given by the long-lived cloud state fully populated ($|c_a|^2 = 1$) and the excited state completely empty ($|c_b|^2 = 0$), the full evolution of the resonant transition is regulated by a single dimensionless parameter (the so-called Landau-Zener parameter)
\begin{equation}
  Z \equiv \left.\frac{(\eta^{(g)})^2}{|g|\gamma}\right|_{\Omega_\textrm{res}},
\end{equation}
where $\gamma \equiv \dot{\Omega}(t)$ is the instantaneous frequency chirp rate at the resonant frequency. The final state of the system after the resonance is in fact described by
\begin{align}
  |c_a|^2 &= e^{-2\pi Z},\\
  |c_b|^2 &= 1-e^{-2\pi Z}.
\end{align}
Hence, for $2\pi Z \gg 1$ the cloud is entirely transferred to the excited state and the resonance is classified as \emph{adiabatic}. Conversely, when $2\pi Z \ll 1$ only a fraction of the cloud is transferred and the resonance is classified as \emph{non-adiabatic}. In a realistic scenario, the states excited by counter-rotating orbits are decaying ($\omega_{n_b\ell_b m_b,\,I}<0$) and the timescale for the decay is usually much shorter than the resonance timescale. This adiabatic elimination of the excited mode, that was first analyzed in \cite{Takahashi2023}, translates into the fact that the fraction of mass that is taken away from the long-lived state falls back into the black hole, bringing to an overall reduction of the cloud mass $M_c$, and a corresponding modification of the mass $M$ and the spin $J$ of the central black hole. For a single transition this is described by
\begin{align}
  \dot{M}_\textrm{c,res} &= \frac{2\Gamma(\eta^{(g)})^2}{\Gamma^2+\Delta^2}M_c\,,\\
  \dot{M}_\textrm{res} &= -2\omega_{n_b\ell_b m_b,\,I}M_{c,b}\,,\\
  \dot{J}_\textrm{res} &= -\frac{2m_b\omega_{n_b\ell_b m_b,\,I}}{\mu}M_{c,b}\,,
\end{align}
where
\begin{align}
  \Gamma &= \omega_{n_b\ell_b m_b,\,I}-\omega_{n_a\ell_a m_a,\,I}\,,\label{eq:gamma_width}\\
  \Delta &= 2(\Omega-\Omega_\textrm{res})\,,\label{eq:delta_width}
\end{align}
and the instantaneous mass in the excited state, $M_{c,b}$, is given by
\begin{equation}
  M_{c,b}(t) = \frac{(\eta^{(g)})^2}{\Gamma^2+\Delta^2}M_c\,.
\end{equation}
This adiabatic elimination is valid under the assumption that $\Gamma^2+\Delta^2\gg (\eta^{(g)})^2$, which we checked to be satisfied for all the resonant transitions that we include in our model.
Resonances in the adiabatic regime, hence, might lead to a complete disruption of the cloud that could happen much before the binary system enters in the frequency band of ground-based detectors. However, the simple picture described above, in which the fate of the resonance depends only on the parameter $Z$ is significantly modified, once one takes into account a series of
mechanisms that can either reduce or enhance the adiabaticity of a given resonance. The most important of these mechanisms comes from the consideration of the backreaction of the resonant transition of the cloud on the orbital evolution of the inspiraling binary. In fact, the transfer of angular momentum between states with different quantum numbers exerts an additional torque on the binary during the transition. The orientation of this torque depends on the difference between the azimuthal quantum numbers, $\Delta m$, of the states, on the orientation (co- or counter- rotating) of the orbit with respect to the cloud, on the sign of the derivatives of the occupation number in each state,
and on the energy difference between the states.
This effect is described by an additional term in the evolution of the binary separation that comes from the conservation of energy and angular momentum in the binary plus cloud system \cite{Baumann2019, Takahashi2023}
\begin{equation}
  P_\textrm{res} = \pm\frac{(m_b-m_a)}{\mu}M^{\frac{1}{2}}(1+q)^{\frac{1}{2}}R^{-\frac{3}{2}}\dot{M}_{c,\textrm{res}}\,,
\end{equation}
with the sign being $+$ for co-rotating orbits and $-$ for counter-rotating  ones.
If the backreaction acts against the natural shrinking of the orbit
($P_\textrm{res}<0$),
the transfer of angular momentum makes the orbit float in the resonance band, thus exciting the transition to the decaying state for a much greater number of cycles increasing the overall adiabaticity of the resonance. Conversely, if the torque acts in the opposite direction
($P_\textrm{res}>0$),
the binary orbit shrinks faster than the vacuum case, driving the system out of the resonance band faster, thus reducing the adiabaticity of the transition. In the inspiral evolution of a binary black hole system with a boson cloud, like the one that we are considering, the system scans increasingly high orbital frequency and undergoes several resonances of both the \emph{sinking} and \emph{floating} kinds.

The resonant history of gravitational atoms for generically eccentric and inclined orbits has been thoroughly investigated in \cite{Tomaselli2024}. Three types of resonances are considered: \emph{hyperfine}, \emph{fine} and \emph{Bohr} resonances, depending on the splitting (see \S\ref{sec:ultralight_bosons}) between the cloud states considered. Due to the very different energy scales corresponding to the distinct splittings between the cloud states, the three types of resonances occur at hierarchically different binary separations. Hyperfine resonances, to which the smallest energy difference correspond, occur at large separations ($R\sim\mathcal{O}(10^4M)$ for $M\mu\sim0.2$) corresponding to the early phases on the binary inspiral. For this reason, the binary can spend a large number of cycles in the resonance band ($\gamma \ll \Omega^2$ and $Z\gg 1$) leading to a strongly adiabatic transition. For the case of a cloud in the $|211\rangle$ that we consider here, the only allowed hyperfine transitions are to the $|21\textrm{-}1\rangle$ and the $|210\rangle$ states, both mediated by the quadrupole $\ell_* =2$. In \cite{Tomaselli2024} it was shown that these resonances, both leading to a floating orbit, can be highly
adiabatic and can lead to a complete disruption of the boson clouds, for realistic values of the binary and cloud parameters, for co-rotating orbital configurations. On the other hand, for a counter-rotating equatorial orbit, these transitions are completely
inhibited (as can be seen from the fact that $\Delta\epsilon<0$ and $\Delta m <0 $)
and leave the cloud intact.
Moreover, in \cite{Takahashi:2021yhy} it has been shewn that the resonance width for the hyperfine transition to the $|21-1\rangle$ state can be highly reduced when the decrease in the cloud's own angular momentum is taken into account, which can vastly inhibit this resonance from happening (especially in nearly equal-mass binaries).

If the cloud is still intact after hyperfine resonances have been met, the binary can arrive in the region ($R\sim\mathcal{O}(10^3M)$  for $M\mu\sim0.2$) where fine resonances, \emph{i.e.} between states with a fine splitting ($\Delta\ell \neq 0$), can be excited. For a cloud in the $|211\rangle$ state, the only allowed fine resonance is to the $|200\rangle$ state.
As for the hyperfine resonance, the fact that $\Delta \epsilon < 0$ and $\Delta m < 0$ completely inhibits the fine resonance for counter-rotating orbits.
However, since $|200\rangle$ has a huge decay width (Eq.~\eqref{eq:gamma_width}), which strongly suppresses the resonant transition, fine resonances are never excited for a cloud in a $|211\rangle$ state, for neither co-rotating nor counter-rotating orbital configuration.
These results are validated numerically in our model.

Finally, in the later stages of the inspiral, the binary encounters the region where Bohr resonances are excited. This happens on typical scales of the cloud size, where, as we will see later, other interaction effects like ionization and accretion of the cloud by the binary drive the inspiral. For the $|211\rangle$ cloud, all the Bohr resonances
(except for the transition to the $|100\rangle$ state which we will analyze later)
are of the sinking type and thus the system is pulled out of the resonance band earlier,
suppressing the adiabaticity of the transition. For this reason, each single Bohr transition transfers a small fraction of the cloud to the decaying states. However, due to the sinking nature of the resonance backreaction on the orbit and their occurrence late in the inspiral, Bohr resonances can still leave detectable signatures in the frequency band of gravitational-wave detectors, whose detectability we want to assess. For this reason, we focus on either counter-rotating orbital configurations, so that the cloud can arrive intact when the binary is in the ionization/accretion/Bohr resonance band, or co-rotating orbital configurations that form at separations that are below the hyperfine resonance radius, so that no adiabatic resonance can disrupt the cloud during the inspiral. The latter can be the result of several of formation mechanisms, \emph{e.g.} dynamical capture \cite{Tomaselli2023, LISA2023}
or the isolated binary evolution channel involving a common envelope phase \cite{Guo2024}.

In Figure \ref{fig:resonances} we display the orbital separations at which transitions from the initial $|211\rangle$ cloud state to states with principal quantum numbers $n_b \geq 3$ (up to $n_b = 14$) are resonantly excited
by a counter-rotating binary.
These resonances
are all of the sinking type and
are driven by dipole $(\ell_* = 1)$ and quadrupole $(\ell_* = 2)$ couplings in a binary with a mass ratio of $q = 0.1$
and $\alpha=0.2$.
Due to the selection rules, each resonance (aside from the $|3\,1\,\textrm{-1}\rangle$ state) forms close doublets. For the dipole coupling, these doublets involve transitions to the $|n00\rangle$ and $|n20\rangle$ states, while for the quadrupole coupling, transitions occur to the $|n1\textrm{-1}\rangle$ and $|n3\textrm{-1}\rangle$ states. Despite their close orbital separations, the second transition within each doublet is heavily suppressed, largely due to the high decay widths of these states. A detailed description of the full set of Bohr resonances that we include in our calculations, and their intensity, can be found in Appendix \ref{app:resonances}. The most significant mass transfers occur at resonances with $n_b = 3$ states, especially for the dipole coupling, with the transition to the $|3\,0\,0\rangle$ state that, alone, can dissipate almost entirely the boson cloud. Notably, for the dipole case,
additional resonances involving $|n_b00\rangle$ states, with $n_b > 3$, contribute non-negligible mass transfer.

The only floating Bohr resonance accessible from the $|211\rangle$ state is the one to the $|100\rangle$ level. Since for this resonance $\Delta \epsilon< 0$ and $\Delta m < 0$, this resonance is only excited by co-rotating binaries. This resonance alone can completely disrupt the cloud, due to its floating nature. However, this transition happens very late in the inspiral, when the cloud has already gone through all the other resonances and all the other interactions of different nature with the binary. This means that the fraction of cloud still surviving up to that point is very small, which, united with the very large decay width of the $|100\rangle$ state, can completely inhibit the float from happening \cite{Tomaselli2024}.

\paragraph*{Ionization}

Transitions between bound and unbound states, analogous to ionization processes in atoms, can occur due to the excitation of the cloud by the binary perturbation when the orbital separation is comparable to the size of the superradiant cloud \cite{Baumann2022}. The backreaction on the orbit of the massive perturber is understood as dynamical friction produced by the gravitational interaction of the falling body with a non-vacuum environment \cite{Annulli2020, Vicente2022, Traykova2023, Tomaselli2023}. Ionization occurs as a partial transfer of the cloud from its starting bound state $|n_b\,\ell_b\,m_b\rangle$ to unbound states $|k;\,\ell\,m\rangle$. Similarly to Eq.~\eqref{eq:bound-bound-mixing}, the mixing between these states is described by a matrix element
\begin{equation}
  \eta_{kb} = \langle k;\ell m|V_*|n_b\ell_bm_b\rangle
  \label{eq:bound-unbound-mixing}
\end{equation}
which satisfies the same selection rules for the angular integral $I_\Omega$ for the discrete transitions in Eqs.~\eqref{eq:selection_1}-\eqref{eq:selection_3}. For this reason, when considering circular equatorial orbits, for which the perturbation is periodical $\phi_* = \Omega t$ (being $\Omega$ the orbital frequency), of all the Fourier components of the matrix element
\begin{equation}
  \eta_{kb} = \sum_{\tilde{g}\in\mathbb{Z}}\eta^{(\tilde{g})}_{kb}e^{-i\tilde{g}\Omega t} \simeq \eta^{({g})}_{kb}e^{-i{g}\Omega t}
\end{equation}
only those oscillating at frequency $g\Omega$ survive, being $g = m - m_b$ for co-rotating orbits and $g = m_b - m$ for counter-rotating ones. This implies that, starting from a cloud that initially only populates a quasi-bound state $|n_b\,l_b\,m_b\rangle$, the total ionization rate can be computed by summing over all unbound states $|k_{(g)};\ell m_b\pm g\rangle$, which yields the rate of variation of the cloud mass
\begin{equation}
  \dot{M_c} = -M_c\sum_{\ell, g}\frac{\mu|\eta^{(g)}_{kb}|^2}{k_{(g)}}\Theta(k_{(g)}^2)
  \label{eq:ionization_mass_rate}
\end{equation}
where $k_{(g)} = \sqrt{2\mu(\omega_{n_b\ell_bm_b}+g\Omega)}$ is the energy of the unbound state (the Heaviside theta $\Theta(k_{(g)}^2)$ guarantees that the sum is over all transitions that actually lead to an unbound state). Similarly one can compute the rate of energy carried away from the system during ionization (the \emph{ionization power})
\begin{equation}
  P_\textrm{ion} = \frac{M_c}{\mu}\sum_{\ell,g}g\Omega\frac{\mu|\eta^{(g)}_{kb}|^2}{k_{(g)}}\Theta(k_{(g)}^2).
  \label{eq:ionization_power}
\end{equation}

\begin{figure*}
  \includegraphics[width = \textwidth]{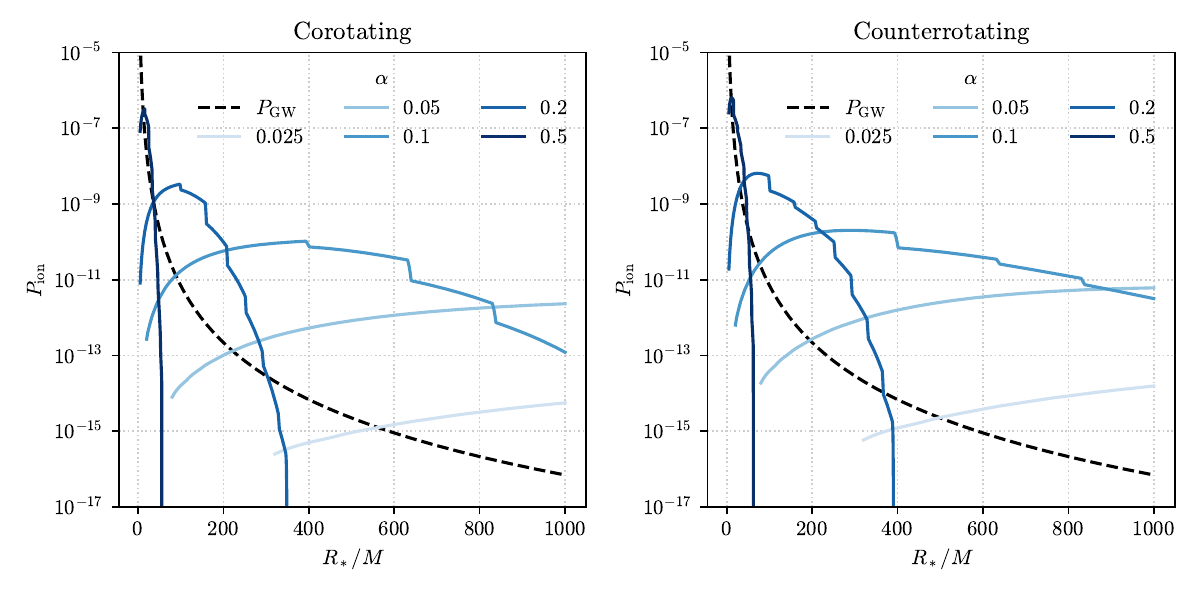}
  \caption{Ionization power in geometrized units for a boson cloud in the $|211\rangle$ state as a function of the orbital separation for different values of $\alpha$ and for our benchmark set of parameters, with $q = 10^{-1}$, $M_c = 0.1M$ and for both corotating (left panel) and counterrotating (right panel) orbital configurations. The dashed line corresponds to the rate of energy loss in the binary system due to the emission of gravitational waves, Eq.~\eqref{eq:p_gw}.}
  \label{fig:ionization_power}
\end{figure*}

An equivalent expression can be derived by considering the particle flux to infinity, as done in \cite{Takahashi:2021yhy, Takahashi:2024fyq}.
As the orbital frequency increases during the inspiral, new ionization channels are opened, with an increasingly high number of overtones $g\Omega$ of the orbital frequency being able to excite transitions to the unbound states. This reflects in the appearance of sharp discontinuities in the ionization power at discrete frequencies (similar to the photoelectric effect \cite{Tomaselli2023})
\begin{equation}
  \Omega^{(g)} = \frac{\alpha^3}{2gMn_b^2}, \qquad g = 1,2,3,\dots,
\end{equation}
with greater values of $\alpha$ corresponding to threshold frequencies that are more squeezed towards the central black hole (reflecting the greater compactness of the cloud). In Fig.~\ref{fig:ionization_power}, we show $P_\textrm{ion}$ for a cloud initially in the $|211\rangle$ state, for both corotating and counterrotating orbital configurations in a binary system with $q = 10^{-1}$
and for different values of $\alpha \in [0.025, 0.5]$. Remarkably, the energy carried away from ionization dominates over the radiation reaction, Eq.~\eqref{eq:p_gw}, due to the emission of gravitational waves (plotted with a dashed line), hinting at strong departures in the evolution of the inspiral compared to the vacuum case, especially true at large orbital separations. Importantly, a scaling relation exist that allows to obtain $P_\textrm{ion}$ for any given set of parameters \cite{Baumann2022}
\begin{align}
  P_\textrm{ion} &= \alpha^5 q^2 \frac{M_c}{M} \mathcal{P}(\alpha^2 R_*/M)\,,\label{eq:P_ion_scaling}\\
  \frac{\dot{M}_c}{M_c} &= \frac{\alpha^3 q^2}{M} \mathcal{R}(\alpha^2 R_*/M)\,\label{eq:M_dot_ion_scaling},
\end{align}
being $\mathcal{P}$ and $\mathcal{R}$ functions that only depend on the initial bound state of the cloud, which we find numerically from Eqs.~\eqref{eq:ionization_mass_rate} and \eqref{eq:ionization_power} for a given set of parameters.

Among the assumptions leading to the expressions in Eqs.~\eqref{eq:ionization_mass_rate} and \eqref{eq:ionization_power}, there is the hypothesis that the system is not in any active bound-bound resonance. This is of course not entirely true for the case of Bohr resonances that occur at the same orbital separations where the ionization is relevant (see Fig.~\ref{fig:interactions}). Nonetheless, in \cite{Tomaselli2024}, where the framework has been extended to account for the extra terms produced by the co-occurrence of ionization and active resonance, it has been demonstrated that such terms are generally negligible for realistic configurations and sets of parameters. For this reason, in this work we consider the two process occurring independently and do not include such modifications.

\begin{figure*}
  \includegraphics[width = \textwidth]{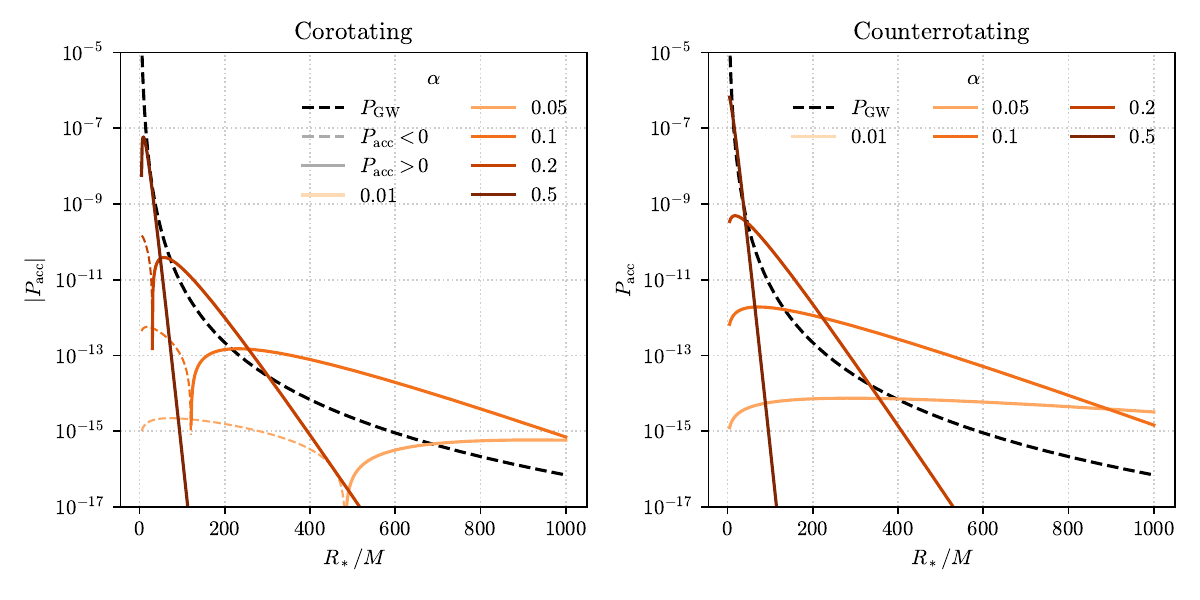}
  \caption{Accretion power in geometrized units for a boson cloud in the $|211\rangle$ state as a function of the orbital separation for different values of $\alpha$ and for our benchmark set of parameters, with $q = 10^{-1}$, $M_c = 0.1M$ and for both corotating (left panel, where we report) and counterrotating (right panel) orbital configurations. The dashed line black line corresponds to the rate of energy loss in the binary system due to the emission of gravitational waves, Eq.~\eqref{eq:p_gw}, while dashed coloured lines in the left panel correspond to orbital separations where $P_\textrm{acc} < 0$, meaning that (in the absence of any other effect) accretion transfers energy from the cloud to the binary.}
  \label{fig:accretion_power}
\end{figure*}

\paragraph*{Accretion}
In \cite{Baumann2022} the problem of accretion of the scalar cloud on the secondary black hole has been considered. Due to the finite size of the companion, part of the scalar field will be absorbed by the secondary black hole, leading to an increase of its mass and thus a backreaction on its orbit. The cloud, on the other hand, although rapidly readjusting to the perturbation and replenishing the underdensity wake trailing behind the companion, experiences a change in its total mass. For the systems that we are considering, the relative velocity between the inspiraling companion and the scalar field can be written as
\begin{equation}
  \delta v = \left|\sqrt{\frac{M}{R}}\mp\frac{m}{\mu R}\right|,
  \label{eq:relative-velocity-acc}
\end{equation}
the sign $\mp$ depending on whether the binary is co-rotating ($-$) or counter-rotating ($+$) with the cloud. Whenever this quantity is within the range
\begin{equation}
  2\pi q \alpha \ll \delta v \ll 1,
  \label{eq:relative_velocity_condition}
\end{equation}
the mass accretion rate on the secondary black hole is approximately independent of the relative velocity and can be expressed as \cite{Baumann2022}
\begin{equation}
  \dot{M}_{*,\textrm{acc}} = A_*\rho(\vec{R}_*),
  \label{eq:accretion-mass-rate}
\end{equation}
being $A_* = 4\pi (2M_*)^2$ the area of the secondary black hole event horizon.
The condition in Eq.~\eqref{eq:relative_velocity_condition} can be violated either at very short orbital separations, when the fluid moves very rapidly from the point of view of the binary companion, or at very large distances when the relative velocity is extremely small. While both conditions can be violated during normal inspirals, for small values of $q$ and $\alpha$ (that we consider in this study) this happens in regions where the cloud is so diluted that accretion would play a negligible role.

Momentum transfer between the cloud and the companion, happening as a consequence of the mass accretion, implies a backreaction on the orbit that must be taken into account when studying the evolution of the orbital separation over time. This corresponds to an additional term in Eq.~\eqref{eq:E_orb_evolution} that is defined as \cite{Baumann2022, Cole2023}
\begin{equation}
  P_\textrm{acc} = \left(\sqrt{\frac{R}{M}}\mp\frac{m}{\alpha}\right)\left(\frac{M}{R}\right)^{3/2}\dot{M}_{{\rm *, acc}}.
\end{equation}
Again, the sign $\mp$ depends on whether the binary is co-rotating ($-$) or counter-rotating ($+$) with the cloud, so that global sign of the accretion term depends on whether the cloud locally rotates faster or slower than the companion (the accretion power can switch sign along the inspiral in the corotating case). Fig.~\ref{fig:accretion_power} displays the accretion power $P_\textrm{acc}$ for a cloud in the $|211\rangle$ state, for both corotating and counterrotating orbital configurations in a binary system with $q = 10^{-1}$
and for different values of $\alpha \in [0.025, 0.5]$. In the corotating case we highlight the orbital separations for which $P_\textrm{acc}$ is negative, leading to a transfer of energy from the cloud to the binary, due to accretion. The amplitude of the energy transfer for accretion is, in general, sub-dominant compared to the one from ionization and is generally below the energy loss due to gravitational waves, except in the region where the cloud is densest, where accretion is maximized.

\subsection{Orbital evolution}
\label{sec:orbit_evolution}
\begin{figure*}
  \includegraphics[width = \textwidth]{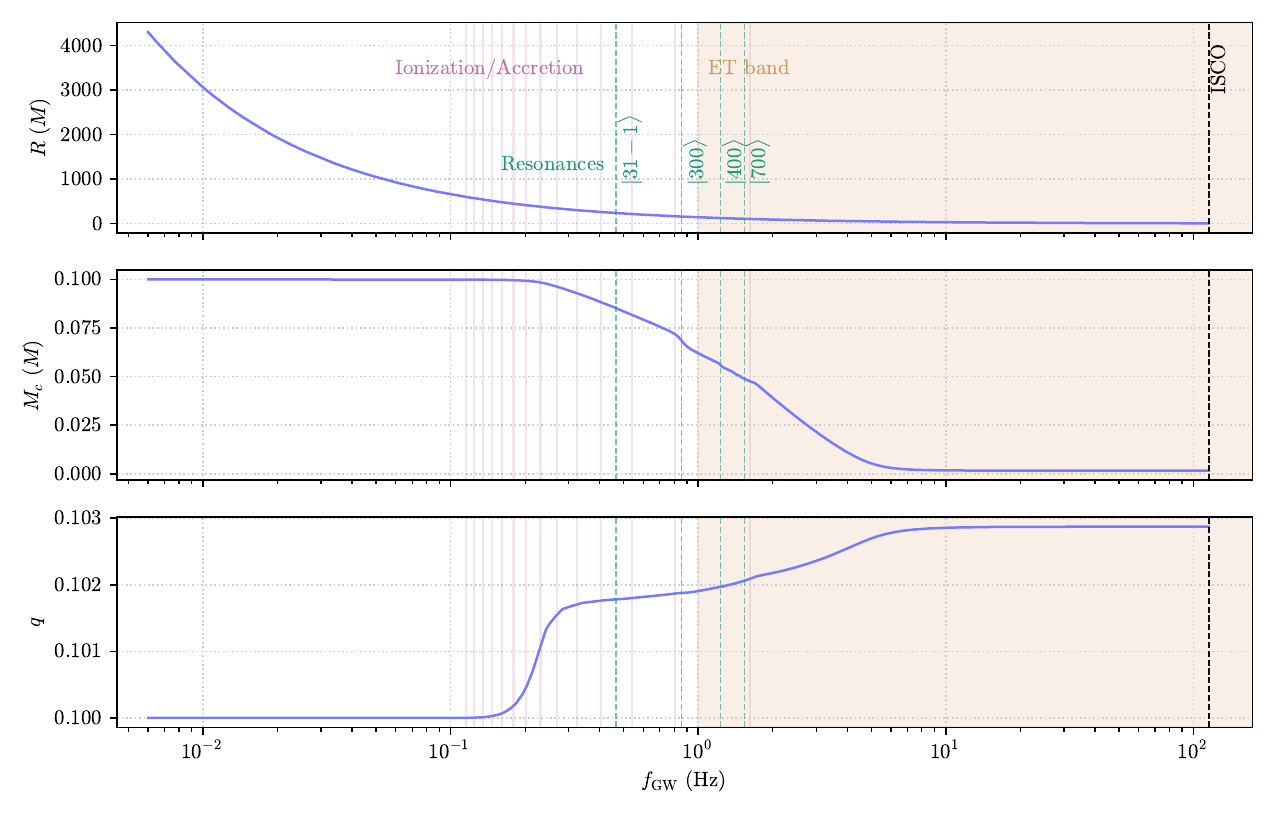}
  \caption{Full evolution, parameterized in terms of the emitted gravitational-wave frequency $f_\textrm{GW}$, for our benchmark system with $M^0 = 40M_\odot$ and $M_*^0=4M_\odot$, so that initial mass ratio is $q \approx 0.1$,
    on a counter-rotating orbit.
    For this particular case, we also fix $\alpha = 0.2$
    and assume that the initial mass of the cloud is 10\% of the central black hole mass. The vertical orange band corresponds to the nominal sensitivity band of ET, while the dashed vertical line represents the ISCO, $f_\textrm{ISCO} \approx 115$ Hz for our choice of parameters. Pink vertical lines, correspond to the sharp discontinuities in the ionization power, Fig.~\ref{fig:ionization_power}, and roughly identify the region where the cloud is prominent. In the same region, Bohr resonances, marked with green vertical lines (we only show
    a selection of the most prominent ones),
  also contribute to cloud depletion.}
  \label{fig:evolution_benchmark}
\end{figure*}

The system of differential equations \eqref{eq:E_orb_evolution}-\eqref{eq:J_evolution} for the state vector of our problem, Eq.~\eqref{eq:dynamical_system}, is integrated numerically starting from a given set of initial conditions
\begin{equation}
  (R^0,\,M_c^0,\,M_*^0,M^0,\,J^0).
\end{equation}
In particular, since we want to analyze the impact on the binary dynamics of all the effects described in \S\ref{sec:binary_interactions}, we start the integration at a large orbital separation (larger than the typical cloud size, so that at the initial time the system is approximately in vacuum).
We fix the mass of the central object to $M^0 = 40M_\odot$ and that of the companion to $M_*^0=4M_\odot$, so that we preserve a small mass ratio $q \approx 0.1$, while producing a gravitational wave signal during inspiral that is above the nominal lower limit of the ET frequency band ($\sim 1$ Hz). Our choice for the black hole masses is made in order to keep the masses to be within a range where such black holes could be of stellar origin while still keeping the mass ratio small since the model should be more accurate the smaller the mass ratio.
It is unclear at the moment exactly up to which mass ratios the model can be trusted. By using $q \approx 0.1$ we might be extrapolating the results, however we will proceed with this choice since it will also allow us to consider corrections to vacuum GR waveform models that are accurate at these mass ratios.
Moreover, for any choice of $\alpha$ and cloud initial mass $M_c$, we fix the angular momentum of the central black hole to the value that saturates the superradiance condition in Eq.~\eqref{eq:superradiant-regime}. The integration is carried out using an implicit Radau scheme \cite{Hairer1993} and is stopped when the binary separation coincides with the Innermost Stable Circular Orbit (ISCO) of the central object, $R_\textrm{ISCO} = 6M$
(for simplicity we neglect the effect of the black hole spin on the ISCO. This does not affect our results since $M_c$ is negligible once $R_\textrm{ISCO} = 6M$ is reached)
, thus covering the entire inspiral phase.

The resulting evolution of the orbital separation, cloud mass and mass ratio is displayed in Fig.~\ref{fig:evolution_benchmark} for our benchmark system in the counter-rotating case and for the choice of parameters, $\alpha = 0.2$, $M_c^0 = 0.1M$. When the system enters the region where the boson cloud peaks, ionization and accretion deplete part of the cloud (almost 50\% of its mass), resulting in a slight increase in the binary mass ratio, as an effect of absorption by the binary companion.

\subsection{Gravitational waveforms}
\label{sec:waveforms}

In this section we describe our methodology to compute realistic gravitational waveforms for the binary coalescence in the presence of a boson cloud, including all the effects described in \S\ref{sec:binary_interactions}.
Similarly to what is done in Refs. \cite{CanevaSantoro2024, Roy2024}, we describe the influence of the non-vacuum environment by introducing a dephasing of the emitted gravitational signal with respect to an inspiral in pure vacuum, ascribable to all the additional energy and angular momentum losses due to the interaction of the binary system with the boson cloud.
More specifically, from the integrated evolution of the binary system, \S\ref{sec:orbit_evolution}, we can compute the orbital frequency of the binary, from which we can obtain the frequency of the emitted gravitational waves as a function of time
\begin{equation}
  f(t) = \frac{1}{\pi}\sqrt{\frac{M(t)+M_*(t)}{R(t)^3}}.
  \label{eq:kepler-third-law}
\end{equation}
From this we can compute the frequency domain strain in both the plus ($+$) and cross ($\times$) polarization,
which can be computed using the stationary
phase approximation \cite{Cutler1994}
\begin{equation}
  \tilde{h}_{+,\times}(f)=\mathcal{A}_{+,\times}(f)e^{i\Phi_{+,\times}(f)},
\end{equation}
where at leading order
\begin{align}
  \mathcal{A}_{+,\times}(f) &= \frac{Q_{+,\times}}{d_L}{\eta}^{1/2} M^{5/6} f^{-7/6},\\
  \Phi_+ & \simeq 2\pi f t(f) - \varphi(f) - \frac{\pi}{4},\\
  \Phi_\times & \simeq \Phi_+ + \frac{\pi}{2}.
\end{align}
Here, $Q_+=(1+\cos^2\iota)/2$ and $Q_\times=\cos\iota$, with $\iota$ the inclination of the binary's orbital angular momentum with respect to the line of sight,
while $d_L$
is the binary luminosity distance from the detector and $\eta$ is the symmetric mass ratio, defined as $\eta = q/(1+q)^2$.
Moreover, $t(f)$ and $\varphi(f)$ correspond to the instant and phase of the gravitational wave signal when it has frequency $f$. Their expression is given by
\begin{align}
  t(f) &= t_c - \int_f^{+\infty} df'\frac{1}{\dot{f}'},\\
  \varphi(f) &= \varphi_c - \int_f^{+\infty} df'\frac{2\pi f'}{\dot{f}'},
\end{align}
being $t_c$ and $\phi_c$ overall time and phase shifts corresponding to the time and phase at coalescence. We compute these integrals numerically for a given system both in presence of the boson cloud ($\Phi_{+,\times}^\textrm{bc}(f)$) and in pure vacuum ($\Phi_{+,\times}^\textrm{vac}(f)$), which allows to compute the waveform dephasing
\begin{equation}
  \delta\Phi(f) = \Phi_{+,\times}^\textrm{bc}(f)-\Phi_{+,\times}^\textrm{vac}(f).
  \label{eq:dephasing}
\end{equation}
When doing so, we fix initial conditions for the evolution of our dynamical system in Eq.~\eqref{eq:dynamical_system} at the entrance of the detector band, in order to make sure that, in the absence of any perturbation due to the cloud presence occurring within the detector band, the two systems (in vacuum and with the cloud) have matching evolutions and thus no dephasing. Estimating the amount of dephasing of a gravitational wave signal from its vacuum counterpart, and the corresponding difference in the number of cycles ($N_\textrm{cyc}^\textrm{vac}-N_\textrm{cyc}^\textrm{bc} = \delta\Phi/2\pi$) from a given reference frequency (usually the lowest detectable frequency of a detector) until the ISCO frequency, can hence be used as a diagnostic of the sensitivity of the instrument to the effects of the presence of a cloud and its ability to modify the in-band signal of the system. As a rule of thumb, we assume that if the signal get dephased by more than 1 cycle (\emph{i.e. $\delta\Phi \sim 2\pi$}) in the considered band this implies detectability.

\begin{figure}
  \includegraphics[width=\columnwidth]{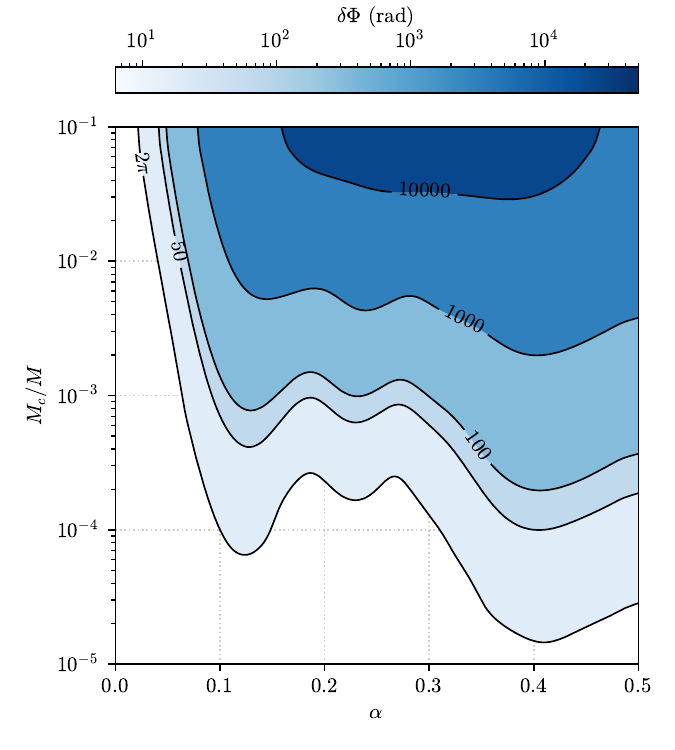}
  \caption{Absolute dephasing in radians from a vacuum waveform accumulated in the frequency band of an ET-like detector computed for our benchmark system and for values of $\alpha\in[0,\,0.5]$ and for cloud masses in the range $M_c\in[10^{-5}M,\,10^{-1}M]$. The level $\delta\Phi = 2\pi$ corresponds to one entire cycle of dephasing, which we assume to correspond to detectability.}
  \label{fig:ET_dephasing}
\end{figure}

\begin{figure}
  \includegraphics[width=\columnwidth]{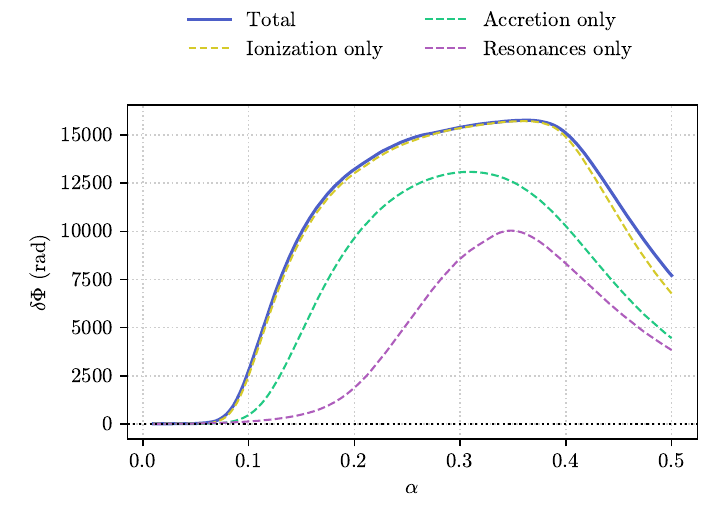}
  \caption{As in Fig.~\ref{fig:ET_dephasing} but fixing $M_c=0.1M$ and by switching off all the interaction effects except one at a time, as reported in the legend. The ionization (yellow dashed line) is the effect that contributes to most of the in-band dephasing.}
  \label{fig:ET_dephasing_effects}
\end{figure}

In Fig.~\ref{fig:ET_dephasing} we show the amount of dephasing in the band of an ET-like detector for our benchmark set of binary parameters and as a function of $\alpha\in[0,\,0.5]$ and $M_c\in[10^{-5},\,10^{-1}]M$. For values of $\alpha > 0.15$ and $M_c > 0.03M$ the amount of dephasing in the ET band can surpass 10000 rad (corresponding to a difference of more than 1500 cycles from the vacuum case). Moreover, looking at the $\delta\Phi = 2\pi$ contour, corresponding to one entire cycle of dephasing, which we assume to be a threshold for detectability, we see sharp features (that are repeated over other contours) that correspond to the entrance in the detector band of ionization thresholds (see Fig.~\ref{fig:ionization_power}) or specific resonances (see Fig.~\ref{fig:resonances}).

To showcase the impact of the different effects of interaction between the binary and the cloud, in Fig.~\ref{fig:ET_dephasing_effects} we report the amount of dephasing with respect to vacuum obtained by switching off all the interaction effects except one at a time. From this analysis, it is clear that the dominant contribution is the dynamical friction on the binary orbit produced through ionization of the boson cloud, $\delta\Phi\approx\delta\Phi_\textrm{ionization}$.
However, a highly non-linear dependence of the dephasing from the individual effects emerges. Intuitively, this behavior can be explained by considering the fact that, as shown in Fig. \ref{fig:ionization_power}, ionization is the driving force of the inspiral when the binary separation matches the cloud size (\emph{i.e.} the same region where accretion and resonances occur). The consequence is that, when ionization is switched on, the binary inspirals much faster through the resonance bands, resulting in a suppression of the resonant transitions and thus an overall smaller signature on the gravitational wave signal of this effect. It is clear that the way the different effects contribute to the total dephasing depends strongly on the cloud parameters and cannot be trivially deduced, as it depends on the intricate interplay between the different effects on the inspiral evolution.

Finally, to more consistently assess the detectability of the cloud with observations from a gravitational wave detectors, we obtain realistic waveforms, by applying our estimate of the dephasing to a pre-existing phenomenological waveform model,
similarly to what is done in Ref. \cite{Roy2024}.
In our case, we adopt the frequency domain phenomenological inspiral-merger-ringdown (IMR) waveform model, \texttt{IMRPhenomXHM} \cite{GarciaQuiros2020} for the inspiral, merger and ringdown of quasi-circular non-precessing black hole binaries, which is publicly available as part of the LIGO Algorithm Library Suite (\texttt{LALSuite}) \cite{LIGO2020_LAL}. Thanks to the calibration of subdominant harmonics to a set of numerical waveforms, this model is suitable for the description of the gravitational signal emitted by highly asymmetric binaries (that can be extrapolated up to case of extreme mass ratios $q\sim10^{-5}$), which is appropriate for the case we are studying.

We assume that the \texttt{IMRPhenomXHM} represents our baseline vacuum waveform $\tilde{h}_{+,\times}^\textrm{vac}(f)$ and, for each mode $(\ell,\,m)$ in the waveform model, we include the dephasing from the presence of the boson cloud by
\begin{equation}
  \tilde{h}_{+,\times}^{\textrm{bc},(\ell,\,m)}(f) = \tilde{h}_{+,\times}^{\textrm{vac},(\ell,\,m)}(f) e^{i\delta\Phi_m(f)},
\end{equation}
where~\cite{GarciaQuiros2020,Mehta:2022pcn}
\begin{equation}
  \delta\Phi_m(f)\approx\frac{m}{2}\delta\Phi\left(\frac{2f}{m}\right),
\end{equation}
with $\delta\Phi(f)$ in Eq.~\eqref{eq:dephasing} corresponding to the dephasing for the dominant $(2,\,2)$ mode. The separate modes are then summed together using spin-weighted spherical harmonics \cite{GarciaQuiros2020} to reconstruct the full waveform  $\tilde{h}_{+,\times}^\textrm{bc}(f)$.
We thus include corrections to the waveform induced by the presence of a non-vacuum environment only in the gravitational-wave phase, neglecting corrections to the gravitational-wave amplitude. This is based on the
fact
that gravitational wave detectors are more sensitive to the evolution of the signal’s phase than its amplitude. Moreover, although formally the correction that we add to the IMR waveform is only valid in the inspiral phase, since the impact of the cloud is mostly relevant in the early inspiral, we can safely assume that the merger-ringdown phases are not affected. This also justifies the use of the simple quadrupole formula, in Eq.~\ref{eq:p_gw}, to model the energy loss by gravitational wave emission in the inspiral phase of the binary evolution (which, as we have shown in Figs.~\ref{fig:ionization_power} and \ref{fig:accretion_power} is sub-dominant compared to the terms arising from the interaction with the cloud).

\begin{figure}
  \includegraphics[width=\columnwidth]{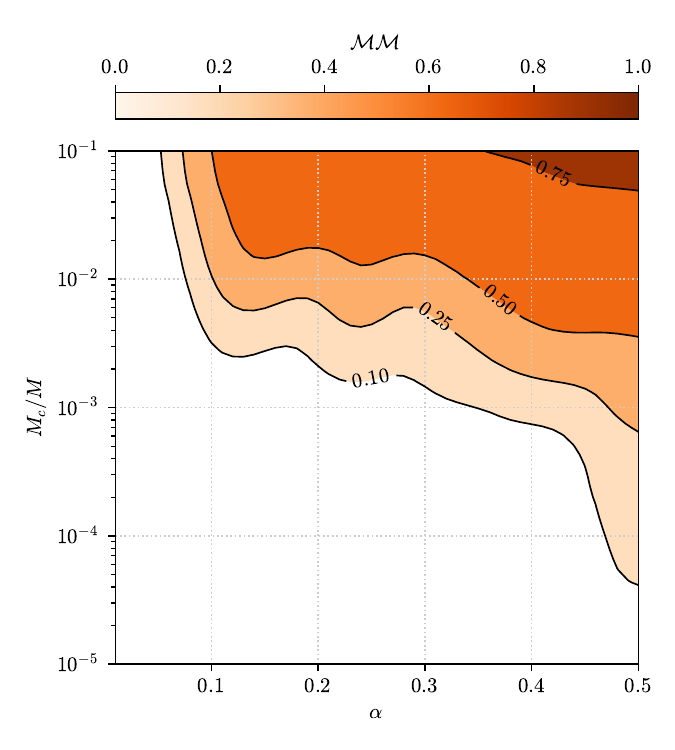}
  \caption{Waveform mismatch for the same benchmark system and the same ranges of parameters considered in Fig.~\ref{fig:ET_dephasing} computed for the ET.}
  \label{fig:ET_mismatch}
\end{figure}

A useful quantity to estimate the difference between a vacuum waveform and one in the presence of a boson cloud is the so called waveform \emph{mismatch}, which is defined as
\begin{equation}
  \mathcal{MM}(h^\textrm{vac},h^\textrm{bc}) = 1-\max_{t_c,\varphi_c} \frac{\langle h^\textrm{vac}|h^\textrm{bc}\rangle}{\sqrt{\langle h^\textrm{vac} | h^\textrm{vac}\rangle\langle h^\textrm{bc} | h^\textrm{bc} \rangle}},
\end{equation}
where the maximization is taken over the overall phase $\varphi_c$ and time $t_c$. The noise-weighted inner product is defined using the detector one-side power spectral density $S_n(f)$ (in our case we use the nominal sensitivity curve for ET presented in \cite{Hild2011}) as
\begin{equation}
  \langle h_1|h_2\rangle = 4\textrm{Re}\int_0^\infty df\frac{\tilde{h}^*_1(f)\tilde{h}_2(f)}{S_n(f)}.
  \label{eq:inner_product}
\end{equation}
The maximization over $\varphi_c$ and $t_c$ can be performed analytically by considering the explicit dependence of the waveforms from $t_c$ and $\varphi_c$ \cite{Owen1996}. In particular, to maximize over $\varphi_c$, one can replace the $\textrm{Re}$ in Eq.~\eqref{eq:inner_product} by the absolute value, corresponding to a rotation of the integral along real axis. On the other hand, a single fast Fourier transform of the integrand function returns the value of $t_c$ that maximizes the inner product. A value of $\mathcal{MM}=0$  implies perfect overlap in the detector band between the two signals, that, due to the maximization procedure, can only differ by a constant phase or time shift. Conversely, a value $\mathcal{MM}=1$ implies zero overlap between the two signals that, thus present a completely different phase evolution in the detector band. In Fig.~\ref{fig:ET_mismatch} we show the resulting mismatch between vacuum waveforms and the ones obtained by introducing the interaction effects with a boson cloud, for our benchmark binary system and different values of $\alpha$ and $M_c$. Interestingly, while both the dephasing and the mismatch roughly peak in the same region of the parameter space, there exist combinations of parameters for which despite a high dephasing, the corresponding mismatch is below 10\%, implying that the in the detector band the two signals can be brought to a high degree of overlap by adding a constant phase shift.

\section{Fisher Matrix Analysis}
\label{sec:fisher_matrix}

The analysis of the dephasing produced in a gravitational wave signal due to the presence of a non-vacuum environment (Fig.~\ref{fig:ET_dephasing}) and the corresponding mismatch (Fig.~\ref{fig:ET_mismatch}) are useful diagnostics to assess the possibility to observe signatures of the presence of the boson cloud, starting from the emitted signal. In order to assess the actual detectability of such effects using current and future detectors, however, a more thorough analysis is needed. A good alternative to a full parameter-estimation analysis, which is very costly especially when one does not have an analytical expression for the waveform modification, as in our case, is to assess the parameter-estimation capabilities of a detector (or a network of detectors) using the Fisher-matrix approximation \cite{Vallisneri2008}. In \S\ref{sec:fisher_matrix_formalism} we give an overview of the formalism behind the Fisher-matrix approximation and we describe how we have implemented it with our waveform model. The results of our analysis are presented in \S\ref{sec:fisher_matrix_results}.

\subsection{Formalism overview}
\label{sec:fisher_matrix_formalism}

The Fisher information matrix for a parametric gravitational waveform model $h(\vec{\theta})$ is defined as \cite{Vallisneri2008}
\begin{equation}
  \mathcal{F}_{ij}(h(\vec{\theta})) = \langle \partial_i h  | \partial_j h \rangle,
  \label{eq:fisher_matrix}
\end{equation}
where $\partial_i h$ is the derivative of the gravitational waveform of interest with respect to the $i$-th parameter. Importantly, the inverse Fisher matrix, $\mathcal{F}^{-1}_{ij}\equiv \mathcal{C}$ is known to coincide with the covariance matrix of the posterior probability distribution for the true source parameters, under the assumption of Gaussian noise and in the high signal-to-noise ratio limit (\emph{i.e.} assuming that the likelihood is Gaussian-like and highly peaked around the true parameters). Despite the fact that some shortcomings, both theoretical and practical, of the use of the Fisher-matrix formalism for gravitational wave analysis have been highlighted over the years \cite{Vallisneri2008, Rodriguez2013}, this method comes with several advantages. First, it allows to get a rough but very fast assessment of the parameters-estimation capabilities (\emph{i.e.} of the uncertainties ) for a given waveform model, avoiding the complications and computational cost of a fully fledged parameter estimation analysis. Additionally, when considering a network of detectors, the Fisher matrices for the diﬀerent components can be added together, regardless of the frequency bands of the single detectors, allowing a straightforward extension of the formalism to multi-band scenarios.

The actual computation of the Fisher matrix in Eq.~\eqref{eq:fisher_matrix} and its inversion, in order to compute $\mathcal{C}$, can be complicated by several factors. For example, in some cases like ours, when the waveform model does not have an analytical expression, the differentiation with respect to the model parameters involved in the definition of the Fisher matrix has to be performed numerically, thus suffering from the limitations of numerical precision. Moreover, the Fisher matrix usually involves derivatives with respect to parameters whose values differ by several orders of magnitude, and it can be nearly degenerate. Both conditions make its numerical inversion potentially problematic. To avoid so, we use the recent and well-tested code \texttt{GWFish} \cite{Dupletsa2023}, which mitigates all such difficulties by combining analytical differentiation with respect to extrinsic waveform parameters (such as $t_c$ and $\varphi_c$) and incorporates routines specifically tailored for the inversion of the Fisher matrix.
Furthermore, \texttt{GWFish} has built-in sensitivity curves for current and future gravitational wave detectors and it is built to work with waveforms from the \texttt{LALSuite}. This makes it particularly suitable for our problem, as we only need to properly implement our numerical estimation for the dephasing into the pre-existing waveform models, as explained in \S\ref{sec:waveforms}.

We consider two sets of injected parameters based on our benchmark binary system as described in \S\ref{sec:orbit_evolution} (for a simpler notation we omit from here on the superscript $0$ to refer to quantities at the entrance of the system in the detector band), and for two different sets of cloud parameters. First, we consider the detectability of the cloud for a system, that we call A, with a value of $\alpha = 0.4$ and initial mass of the cloud of $1\%$ of the central black hole mass. This system falls in the region corresponding to a total in-band dephasing for ET between $100$ and $1000$ full cycles
(see Figure \ref{fig:ET_dephasing}). Then, we chose a system, B, with $\alpha = 0.1$ and $M_c=0.1\%M$, for which the effects related to the presence of the cloud are supposed to be much weaker. Since, we are considering different values of the boson mass $\mu$ in the two cases, the primary spin $a_1$ of the central black hole is different in systems A and B and chosen to always saturate the superradiance condition in Eq.~\eqref{eq:superradiant-regime}. We perform for both systems a Fisher matrix analysis, in which we consider variations of the following set of parameters:
\begin{equation}
  (\alpha,\,M_c,\,M,\,M_*,\,a_1,\,a_2,\,t_c,\,\varphi_c).
  \label{eq:fisher_parameters}
\end{equation}
Other intrinsic parameters, like the tilt angle between the black holes' spins, as well as extrinsic parameters like the sky localization, the luminosity distance
and the inclination angle
of the binary are considered to be fixed in our analysis. A detailed overview of the injected parameters for both cases is reported in Table \ref{tab:fisher_matrix}.

\begin{table*}[]
  \setlength{\tabcolsep}{20pt}
  \renewcommand{\arraystretch}{1.6}
  \begin{tabular}{l|lcccc}
    \hline
    & Parameter (unit)                                      &  Model & Injected          & $\sigma$ (ET)      & $\sigma$ (LIGO + VIRGO ) \\ \hline
    \multirow{4}{*}{
      \begin{sideways}Cloud
    \end{sideways}}  & \multirow{2}{*}{$\alpha$}          & (A) & 0.4                  & $8.3\times10^{-6}$ & $8.1\times10^{-4}$                         \\
    &                                                       & (B) & 0.1                  & $2.9\times10^{-4}$ & 0.29                    \\ \cline{2-6}
    & \multirow{2}{*}{$M_c$ ($M$)}                          & (A) & 0.01                 & $2.5\times10^{-6}$ & $7.6\times10^{-5}$                          \\
    &                                                       & (B) & 0.001                & $4.8\times10^{-5}$ & 0.32                     \\ \hline
    \multirow{8}{*}{
      \begin{sideways}Binary
    \end{sideways}} & \multirow{2}{*}{$M$ ($M_\odot$)}   & (A) & \multirow{2}{*}{40}  & $2.9\times10^{-4}$ & 0.012                         \\
    &                                                       & (B) &                      & 0.119              & 5.4                      \\ \cline{2-6}
    & \multirow{2}{*}{$M_*$ ($M_\odot$)}                    & (A) & \multirow{2}{*}{4}   & $1.4\times10^{-5}$ &  0.0017                    \\
    &                                                       & (B) &                      & 0.002              & 0.34                    \\ \cline{2-6}
    & \multirow{2}{*}{$a_1$}                                & (A) & 0.6                  & $1.4\times10^{-5}$ & 0.0070                        \\
    &                                                       & (B) & 0.198                & 0.046              & 0.093                    \\ \cline{2-6}
    & \multirow{2}{*}{$a_2$}                                & (A) & \multirow{2}{*}{0.2} & 0.004              & 0.019                 \\
    &                                                       & (B) &                      & 0.0022             & 0.079                    \\ \hline
    \multirow{4}{*}{
      \begin{sideways}Extrinsic
    \end{sideways}}  & $\iota$                                   &     & $\pi/3$ \\
    & $z$                                   &     & 0.5 \\
    & $d_L$ (Gpc)                                           &     & 2.9 \\  \cline{2-6}
    & \multirow{2}{*}{SNR}                                  & (A) &                      &  90                & 13                   \\
    &                                                       & (B) &                      &  87                & 12                   \\ \hline
    & $N_\textrm{cycles}$                   &  &                      &  2690              & 260 \\ \cline{2-6}
    & \multirow{2}{*}{$\delta\Phi$}                         & (A) &                      &  4580             &  8.6                  \\
    &                                                       & (B) &                      &  48            &  0.85                  \\ \hline
  \end{tabular}
  \caption{Summary of the Fisher matrix analysis results for systems A and B. The injected values of the cloud and binary parameters are listed alongside their resulting uncertainties, $\sigma$, from the Fisher matrix analysis for ET and LIGO+VIRGO. For the cloud parameters ($\alpha$ and $M_c$), ET provides significantly tighter constraints, particularly in system B, where current-generation detectors fail to resolve the presence of the cloud. The uncertainties in the binary parameters ($M$, $M_*$) and dimensionless spin components ($a_1$, $a_2$) also show considerable improvement with ET, underscoring the influence of the environment on the precision of parameter estimation. Here we also report the details of the parameters used for the waveform generation (and not included in the Fisher matrix analysis) including the inclination ($\iota$) of the binary’s orbital angular momentum with respect to the line of sight,  redshift ($z$), luminosity distance ($d_L$) and signal-to-noise ratio (SNR). Note that all reported masses are redshifted masses.
  Moreover, we report the in-band phase cycles ($N_{\textrm{cycles}}$) for the considered binary in the vacuum case and the corresponding dephasing ($\delta\Phi$) from vacuum resulting from the presence of the cloud.}
  \label{tab:fisher_matrix}
\end{table*}

\subsection{Results}
\label{sec:fisher_matrix_results}

\begin{figure*}
  \includegraphics[width=\textwidth]{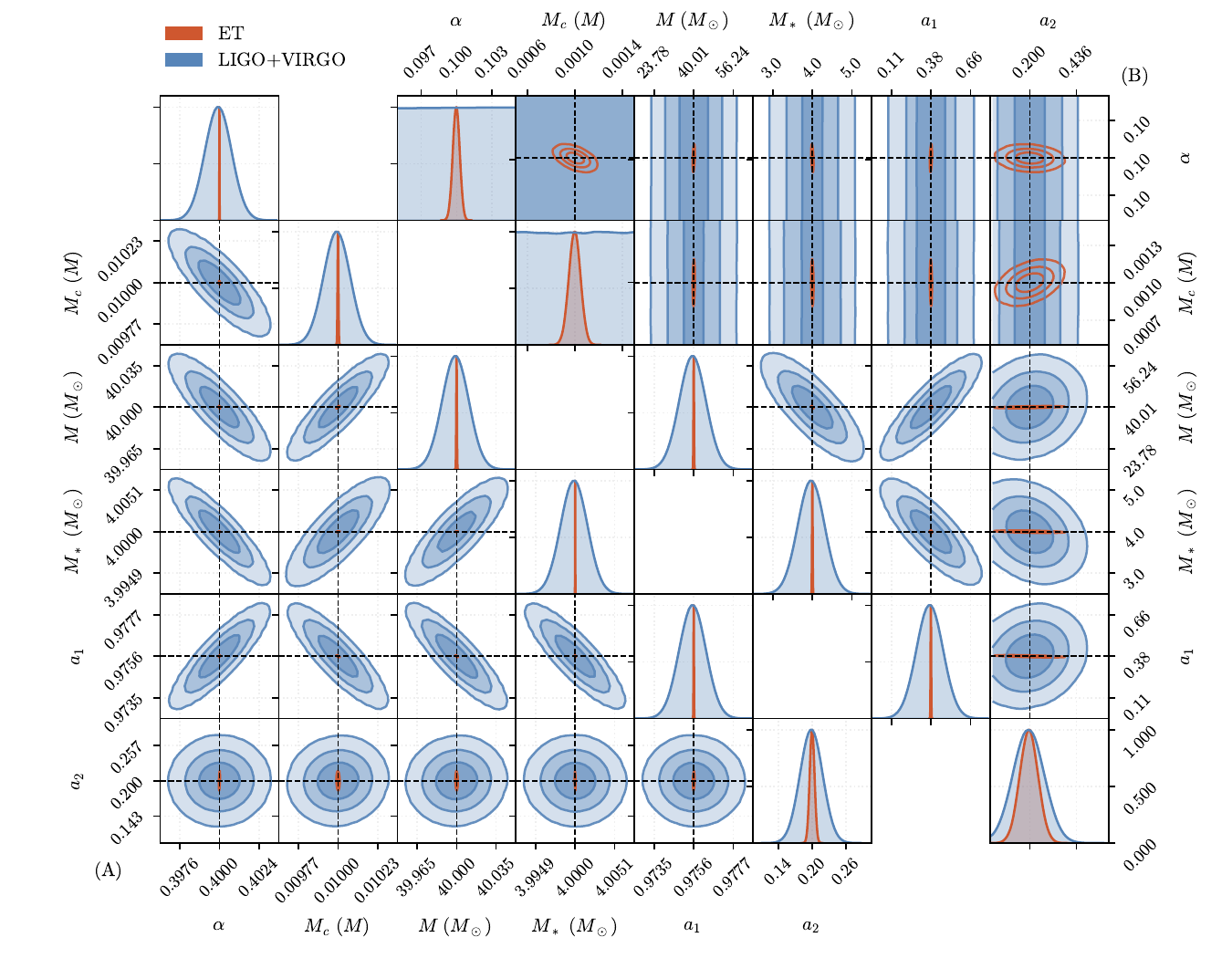}
  \caption{Posteriors from the Fisher matrix analyses for the systems A (bottom-left corner plot) and B (top-right corner plot) with superimposed contours for ET (red) and LIGO+VIRGO estimations. While for the system A both ET and LIGO+VIRGO are able to constrain the parameters of the cloud with a factor $\sim10$ improvement on the constraining capabilities of ET vs. current-generation detectors, in case B LIGO+VIRGO observations are unable to successfully detect the presence of the cloud and allow a direct constraint of its parameters. This also reflects in a significantly poorer constraining power on the binary parameters (e.g. the masses), in agreement with the idea that characterizing effects related to the astrophysical environment of black hole binaries can improve the constraining capabilities of coalescing binaries.
  }
  \label{fig:ET_LIGO_VIRRGO_fisher_matrix_corner}
\end{figure*}

Here we describe the results of our Fisher matrix analysis on the considered systems A and B. In Fig.~\ref{fig:ET_LIGO_VIRRGO_fisher_matrix_corner} we report the corner plots of the reconstructed posterior for the two system assuming observations with either ET
in a triangular configuration
(red contours) and LIGO+VIRGO
at design sensitivity (we exclude the extrinsic parameters, like the phase and time of coalescence, from the plots even though they are included in the Fisher Matrix analysis).
For system A, both LIGO+VIRGO and ET can constrain the parameters of the binary and the cloud environment, though ET achieves approximately a hundredfold improvement in precision. In contrast, for system B, LIGO+VIRGO observations cannot effectively detect the environmental signature (with uncertainties on the cloud parameters that correspond to a $\gtrsim100\%$ relative error), leading to broader and less informative posterior distribution for the binary parameters, such as the component masses and spin parameters. The significant improvement with ET in this case highlights the synergy between precise environmental modeling and advanced detector capabilities, supporting the idea that including astrophysical environmental effects in waveform models can critically enhance the constraining power for binary parameters \cite{Cole2023}.
Table \ref{tab:fisher_matrix} complements Figure \ref{fig:ET_LIGO_VIRRGO_fisher_matrix_corner}, quantifying the results from our Fisher matrix analysis. We list the injected values for both cloud and binary parameters, as well as the resulting uncertainties from the Fisher matrix analysis for systems A and B, observed with ET and LIGO+VIRGO. Moreover, we list the details of the parameters used for the waveform generation (and not included in the Fisher matrix analysis) including the systems redshift and luminosity distance, alongside the corresponding signal-to-noise ratio. Finally, we report the in-band number of cycles for the two detector networks and the resulting dephasing from vacuum, resulting from the presence of the cloud, as defined in Eq.~\eqref{eq:dephasing}.

\begin{figure*}
  \includegraphics[width=\textwidth]{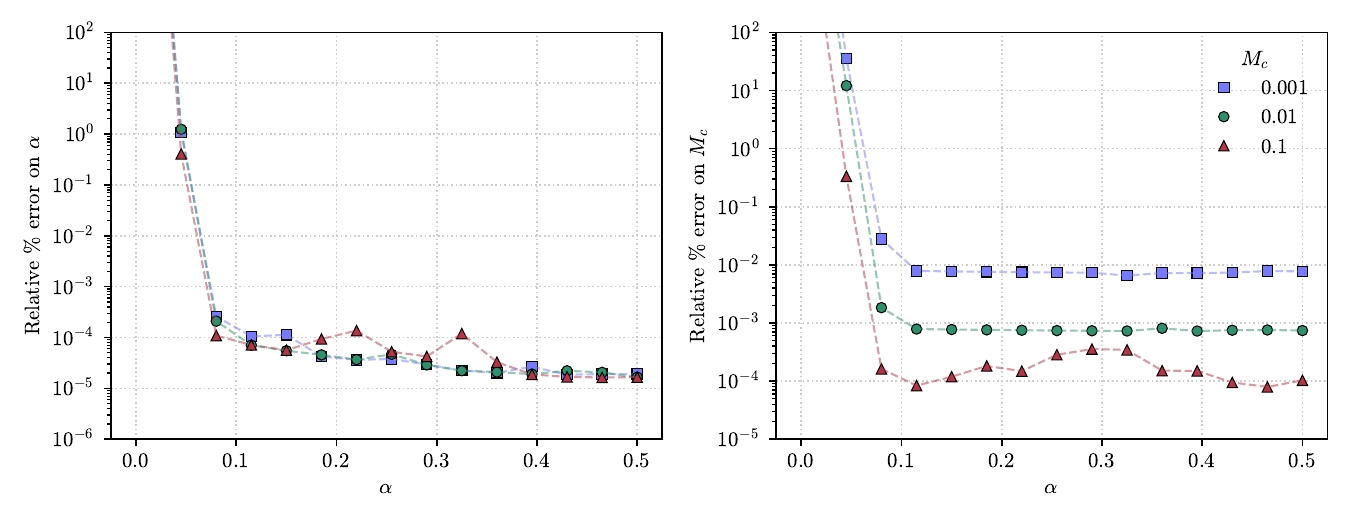}
  \caption{Results of Fisher matrix analyses
    for ET showing the relative uncertainties on the cloud parameters $\alpha$ (left panel) and $M_c$ (right panel), evaluated for 15 equally spaced values of $\alpha \in [0, 0.5]$ and three cloud masses: $M_c = 10^{-1}M$ (red triangles), $10^{-2}M$ (green dots), and $10^{-3}M$ (blue squares). The relative error on $\alpha$ is consistent across the different cloud masses, decreasing to $\sim10^{-4}\%$
    for $\alpha \gtrsim 0.2$ but degrading significantly as $\alpha \to 0$, particularly below $\alpha \lesssim 0.1$. For the cloud mass $M_c$, heavier clouds ($M_c = 0.1M$, red) are better constrained, with relative errors as low as $\sim10^{-4}\%$, while smaller clouds ($M_c = 0.001M$, blue) exhibit larger uncertainties, reaching $\sim10^{-2}\%$
    or more. Cases with relative errors exceeding 100\% of the injected values are considered non-conclusive detections, such as $\alpha \lesssim 0.045$ for $M_c = 0.001M$ and $\alpha \lesssim 0.025$ for $M_c = 0.01M$.
  }
  \label{fig:ET_fisher_matrix_detectability}
\end{figure*}

In Fig.~\ref{fig:ET_fisher_matrix_detectability} we also report the results
for ET
of Fisher matrix analyses performed on our benchmark systems (by injecting the same binary and extrinsic parameters as reported in Table \ref{tab:fisher_matrix}) and considering 15 equally spaced values of $\alpha\in[0,\,0.5]$ and three values of $M_c=10^{-1}M$, $10^{-2}M$ and $10^{-3}M$. In particular, for each system we show the relative posterior uncertainty estimated through the Fisher matrix on the parameters $\alpha$ and $M_c$ (we left as free parameters in our analyses all those appearing in Eq.~\eqref{eq:fisher_parameters}). Interestingly, the precision of the constraints on $\alpha$ (left panel) remains consistent across the three cloud masses considered, with a relative error on $\alpha$ that decreases down to order $\sim 10^{-4}\%$
for $\alpha \gtrsim0.2$ and gets gradually degraded for values of $\alpha$ approaching 0. On the other hand, the relative errors on the cloud mass (right panel) show how a heavier cloud can be characterized much better, with relative errors that can be as low as $\sim10^{-4}\%$ for $M_c =0.1M$,
while it can still be constrained at order $10^{-2}\%$ for a cloud mass as low as $M_c=0.001M$. Whenever the relative error exceeds 100\% of its injected values we consider this case as a non-conclusive detection of the cloud. For example, this happens for values of $\alpha \lesssim 0.045$ for $M_c = 0.001M$ and $\alpha \lesssim 0.025$ for $M_c = 0.01M$.
The curves with $M_c = 0.1M$ exhibit some ``bumps'' corresponding to a slight loss in precision of the cloud's parameters estimation at around $\alpha\sim0.3$. Interestingly, this region corresponds to where the dephasing from resonances alone peaks (see Fig.~\ref{fig:ET_dephasing_effects}). To further understand these features, we have repeated the Fisher matrix study without including the resonance contributions in the waveform. The results, which we show in Appendix~\ref{app:detectabilty-no-resonances}, confirm that the overall trends in parameter precision remain consistent with those discussed earlier. However, the curves are significantly smoother, and the previously observed bumps at specific values of $\alpha$ disappear entirely. Although somewhat counterintuitive, we interpret the loss of precision caused by resonances as possibly due to the fact that all resonances considered are of the sinking type. Their effect might reduce the total number of waveform cycles in band, thereby reducing the amount of information available to constrain the cloud parameters.

\begin{figure*}
  \includegraphics[width=\textwidth]{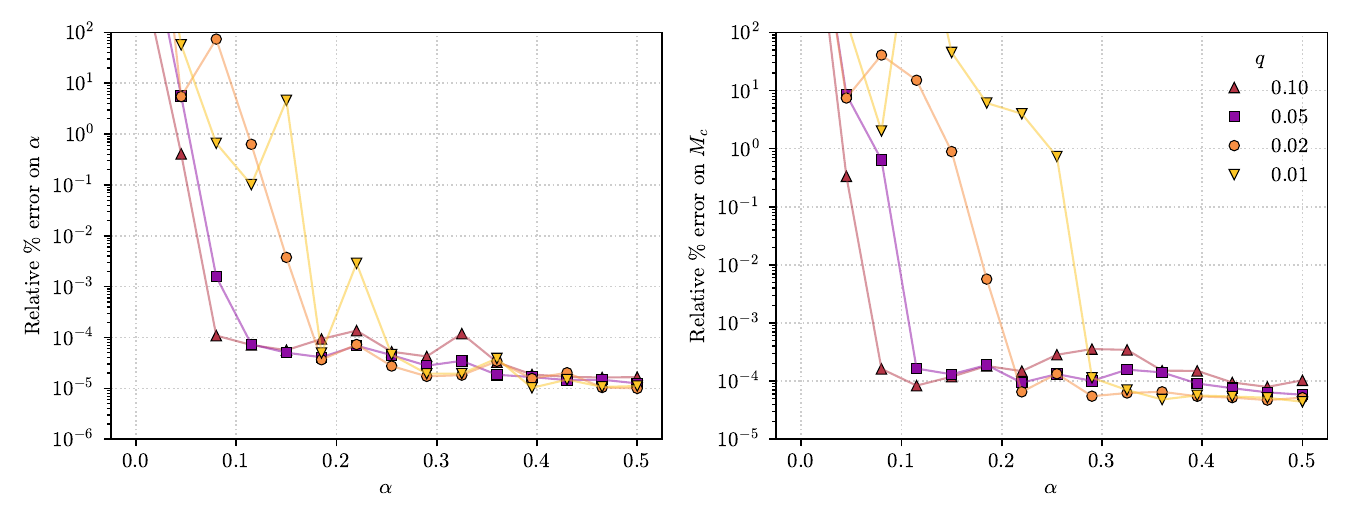}
  \caption{Relative percentage errors on the scalar cloud parameters, $\alpha$ (left panel) and the cloud mass $M_c$ (right panel) as functions of $\alpha$, for different binary mass ratios $q$. All curves are computed at fixed cloud mass $M_c = 0.1M$ and fixed luminosity distance (implying that the red triangles in this plot match exactly the results presented in Fig.~\ref{fig:ET_fisher_matrix_detectability} for $M_c=0.1M$), with the mass ratio varied by changing the primary mass while keeping the companion mass fixed at $M_*=4\,M_\odot$. For lower mass ratios, the inspiral occurs predominantly outside the ET frequency band, leading to a loss in precision for small $\alpha$. High precision is recovered only for larger $\alpha$, when interaction effects fall back into the detector band (\emph{e.g.} $\alpha \gtrsim 0.1$ for $q = 0.05$, and $\alpha \gtrsim 0.3$ for $q = 0.01$). The appearance of peaks, especially in the lowest-$q$ curves, may indicate specific resonances or ionization discontinuities entering/exiting the detector band. Note that as the luminosity distance is fixed, the signal-to-noise ratio (SNR) varies across curves, which can also contribute to the observed loss in parameter precision at smaller $q$.}
  \label{fig:ET_fisher_matrix_detectability_diffferent_qs}
\end{figure*}

In Fig.~\ref{fig:ET_fisher_matrix_detectability_diffferent_qs}, we complement the above analysis by exploring how the detectability of the cloud parameters depends on the binary mass ratio, while keeping fixed the cloud mass to $M_c = 0.1M$. Specifically, we consider four different values of the mass ratio $q = \{0.10,\, 0.05,\, 0.02,\, 0.01\}$, which we obtain by varying the primary mass (while keeping the companion mass fixed at $M_*=4M_\odot$), and compute the relative uncertainties on $\alpha$ and $M_c$ across 15 equally spaced values of $\alpha\in[0,\,0.5]$, by repeating our Fisher Matrix analysis for such combinations of parameters. As evident from the left panel, smaller mass ratios lead to a degradation in the constraints on $\alpha$ at low $\alpha$ values. This can be attributed to two effects: first, the inspiral phase increasingly occurs outside the sensitivity band of ET, as $q$ gets smaller, pushing interaction-induced modifications to the waveform outside the ET band; and second, since we keep the luminosity distance fixed, the SNR naturally decreases for smaller mass ratios (going from $\textrm{SNR} \sim 87$ for $q=0.1$ to $\textrm{SNR} \sim 70$ for $q = 0.01$, in the vacuum case). Nonetheless, high precision (relative errors $\lesssim 10^{-3}\%$) is recovered for larger values of $\alpha$ (e.g., $\alpha \gtrsim 0.1$ for $q = 0.05$ and $\alpha \gtrsim 0.3$ for $q = 0.01$), where the cloud’s effect becomes strong and re-enters the detector band. Similarly, the right panel shows that the cloud mass $M_c$ can still be constrained with excellent accuracy in this regime, with the precision improving slightly as $q$ decreases. Sharp peaks visible in the lowest-$q$ curves are likely associated with resonances or ionization features moving in and out of the detector band, although a denser sampling in $\alpha$ would be required to characterize this effect more robustly.

\section{Discussion and Conclusions}
\label{sec:discussion}

The advent of gravitational wave astronomy has opened unprecedented opportunities to explore fundamental physics \cite{Barack2019, Sathyaprakash2019, Bertone2020} and probe the environments surrounding compact objects like black holes \cite{Cardoso2020, Cardoso2022, Cole2023, Zwick2023, Eda2013, Kavanagh2020, Hannuksela2020, Yunes2011, Derdzinski2021}. Among the most intriguing possibilities is the detection of ultralight bosons, proposed as candidates for beyond-Standard-Model particles \cite{Peccei1977, Arvanitaki2010} and dark matter \cite{Hui2017, Hui2021}. Through the process of superradiance \cite{Brito2015}, these bosons can form macroscopic clouds around rotating black holes, leading to distinctive interactions in coalescing binary systems \cite{Baumann2020, Baumann2022,Takahashi2023, Tomaselli2023,Tomaselli2024}. This provides a natural laboratory to detect these elusive particles and constrain their properties. In this context, next-generation gravitational wave observatories, such as LISA \cite{Colpi2024} and ET \cite{Maggiore2020}, designed to explore lower gravitational wave frequencies compared to current-generation observatories and to achieve at the same time improved sensitivity, promise to transform our ability to successfully detect the presence of boson clouds and estimate their physical parameters.

In this work, we investigated the detectability of ultralight boson clouds formed through black hole superradiance in compact binary systems using next-generation ground-based gravitational wave observatories, like ET. Our numerical framework includes a comprehensive characterization of all the relevant interaction effects between the boson cloud and the binary system — resonances, accretion, and ionization — and their interplay, being based on state-of-the-art modeling on the theoretical side \cite{Baumann2020, Baumann2022,Takahashi2023, Tomaselli2023,Tomaselli2024}.
Our results demonstrate that the presence of a boson cloud can induce significant dephasing in the gravitational signal, leaving a distinctive imprint that could be detected with next-generation observatories like ET. More specifically, we found that the dephasing caused by the cloud can reach detectable levels
for a wide range of cloud and boson masses. Additionally, using the Fisher Matrix information formalism, we showed that enhanced sensitivity with ET will allow not only robust detection of these imprints but also a precise estimation of cloud parameters, such as its mass and the gravitational fine structure constant, significantly improving upon current capabilities.
In this context, our work represents an advancement over previous detectability studies with LISA, like \cite{Cole2023}, not only because it extends their scope to ground-based detectors, but also in terms of theoretical modeling. Such past studies, in fact, typically focused on individual effects in isolation or only considered a subset of them. Our model, on the other hand, considers the full interplay between all relevant cloud-binary interaction effects and also allows for assessing the impact on the observational signatures of each one, individually. Moreover, while previous detectability studies focused on assessing the parameter estimation capabilities only for a benchmark system, here we presented a systematic exploration of the parameter space, which was enabled by the use of the Fisher Matrix formalism. Finally, we build on top of accurate IMR vacuum waveforms (compared to the Newtonian order waveforms used \emph{e.g.} in \cite{Cole2023}), which represents a significantly more realistic and precise modeling of the gravitational signal allowing to properly characterize the response of current and future ground-based detectors.

On the other hand, it is also important to highlight the current limitations of our approach, that leave room for future improvements. We have restricted our analysis to circular equatorial orbits, which, apart from the great reduction in complexity of the problem, also maximize the interaction effects and the corresponding gravitational wave signature. Realistic binary systems, with non-circular and inclined orbits, can be characterized by additional dynamical features and a much richer interaction phenomenology with boson clouds \cite{Baumann2022, Tomaselli2024, Boskovic2024,Tomaselli2024b}. Most strikingly, generic orbital configurations can lead to an early disappearance of the cloud \cite{Takahashi2023, Tomaselli2024}, which suggests that the chances of directly detecting ultralight bosons when the considered binary systems enter the frequency band of ground-based detectors can be vastly reduced. This is strictly connected to the other main limitation of our current approach: the physical properties of the boson cloud, fixed at the moment of time when the binary system enters the detector band, were treated as free parameters.
The main conclusion of our work is that \emph{if} the cloud can survive up to entering the ET band, then its signatures are potentially detectable in part of the parameter space. However, we did not model explicitly the formation history of the cloud or its potential dissipation due to gravitational wave emission or to the tidal perturbation from the companion in the binary system at earlier times. Exploring the full history of the cloud since its formation would be highly important to understand the likelihood of observing such systems.

Additionally, all the effects we considered here were computed using a Newtonian approximation. Work towards modelling the impact of boson clouds in the strong field regime has recently started to be done employing a small mass ratio approximation \cite{Duque:2023seg,Brito2023,Dyson2025}. Whether these calculations can be extrapolated to the mass ratios we considered in this work ($q\sim 0.1$) remains to be fully understood.
The limitation of the model we employed to small mass-ratios is also an important set back of our work. Extending our work to nearly equal-mass ratios would require a different treatment of the system, especially once the cloud surrounds both black holes. Understanding how to treat nearly equal-mass ratios is also crucial in order to know up to which mass ratios can our model be trusted. Moreover, we only considered a cloud in the $|211\rangle$ state. As we already emphasized, a realistic treatment of the evolution of the cloud would also need to include higher-order unstable modes, especially for large values of $\alpha$.
We also did not include finite-size effects from the cloud that could potentially be relevant, such as the impact of non-trivial tidal Love numbers~\cite{Baumann2019,Baumann2020,DeLuca:2021ite,DeLuca:2022xlz,Chia:2023tle,Arana:2024kaz} or the intrinsic multipole moments of the cloud~\cite{Baumann2019,Baumann2020}. Although finite-size effects are likely to be subdominant with respect to ionization in the regime where the secondary object is moving inside the cloud, a fully accurate waveform in the whole inspiral will probably need to also take these effects into account.

Moreover,
on the data analysis side,
our current parameter estimation is based on the Fisher Matrix analysis. While this technique is universally regarded as a powerful tool to forecast the precision that will be reached by a specific future experiment \cite{Vallisneri2008}, it does not allow to assess the possible emergence of biases in the parameter estimation of a given model nor it does account for non-Gaussian posteriors which can lead to a great mis-estimation of the parameters' uncertainties, especially for large and highly degenerate parameter spaces. Additionally, the hypothesis that the inverse Fisher-Matrix can be used as an estimator of the covariance matrix only applies in the high-SNR case. At low SNR, this assumption breaks down, leading to unreliable estimates of parameter uncertainties, which is especially relevant for the constraints that we have shown for the LIGO+VIRGO case, that must thus be considered with care.

We plan to address these shortcomings in future works. The first natural step would be to perform a full Bayesian parameter estimation to attenuate the limitations imposed by the Fisher Matrix approach. Moreover, our numerical framework can be extended to include non-circular and inclined orbital configurations, leveraging recent theoretical modeling of these systems \cite{Tomaselli2023, Tomaselli2024,Boskovic2024,Tomaselli2024b}. This would also involve a more detailed modeling of the boson cloud formation and evolution, including dissipation effects that we have neglected.
Performing a full exploration of the parameter space of possible binary black hole systems,
cloud states, as well as their history since formation
would also be important in order to understand the exact range of boson masses detectable with future gravitational wave detectors.
Moreover, extending the analysis to include massive vector bosons \cite{Dolan2018,Cao:2024wby} would expand the scope of the study to a broader class of ultralight bosonic fields. Finally, an interesting extension of the current work would be to consider how a multi-band scenario \cite{Sesana:2016ljz}, exploiting the synergy of future ground-based observatories and space-based observatories, such as LISA, could enable more precise parameter estimation and enhance the detectability for boson clouds around rotating black holes.

\acknowledgments
We are grateful to Rafael Porto, Mateja Bo{\v{s}}kovi{\'c, Matthias Koschnitzke, Rodrigo Vicente and Thomas Spieksma for useful comments on the final draft of this manuscript. R.B. acknowledges financial support provided by FCT – Fundação para a Ciência e a Tecnologia, I.P., under the Scientific Employment Stimulus -- Individual Call -- Grant No. \href{https://doi.org/10.54499/2020.00470.CEECIND/CP1587/CT0010}{2020.00470.CEECIND}, the Project No. \href{https://doi.org/10.54499/2022.01324.PTDC}{2022.01324.PTDC} and the Project ``GravNewFields'' funded under the ERC-Portugal program. RDM acknowledges support from the  grant PID2021-122938NB-I00 funded by MCIN/AEI/10.13039/501100011033 and by ``ERDF A way of making Europe'', from the Consejeria de Educación de la Junta de Castilla y León and the European Social Fund. Moreover, RDM acknowledges the  Network in Gravity (\href{https://sites.google.com/view/gravnet/}{GravNet}) for financing a Short Term Scientific Mission that made this work possibile.

  \bibliographystyle{apsrev4-2-author-truncate}
  \bibliography{biblio}

  \appendix

  \vspace{2cm}

  \section{Details on resonant transition for our benchmark system}
  \label{app:resonances}

  \begin{table*}[!t]
    \setlength{\tabcolsep}{12pt}
    \renewcommand{\arraystretch}{1.4}
    \begin{tabular}{lccccccr}
      \hline
      State  & $\ell_*$ & $m_*$ & Type & $Z$ & $\eta^{(g)}$ & $R_\textrm{res}$ & $\Delta M_c$ (\%)\\ \hline
      $|3\,1\,\textrm{-1}\rangle$ & 2 & -2 & Sinking & $28$ & $3.3 \times 10^{-6}$ & $236.7$ & $-7.8$ \\
      $|4\,1\,\textrm{-1}\rangle$ & 2 & -2 & Sinking & $4.1$ & $2.2 \times 10^{-6}$ & $193.3$ & $-1.6$ \\
      $|5\,1\,\textrm{-1}\rangle$ & 2 & -2 & Sinking & $1.3$ & $1.5 \times 10^{-6}$ & $179.4$ & $-0.61$ \\
      $|6\,1\,\textrm{-1}\rangle$ & 2 & -2 & Sinking & $0.58$ & $1.1 \times 10^{-6}$ & $172.7$ & $-0.3$ \\
      $|7\,1\,\textrm{-1}\rangle$ & 2 & -2 & Sinking & $0.31$ & $8.8 \times 10^{-7}$ & $169.1$ & $-0.17$ \\
      $|8\,1\,\textrm{-1}\rangle$ & 2 & -2 & Sinking & $0.19$ & $7.1 \times 10^{-7}$ & $166.8$ & $-0.11$ \\
      $|9\,1\,\textrm{-1}\rangle$ & 2 & -2 & Sinking & $0.12$ & $5.9 \times 10^{-7}$ & $165.2$ & $-0.071$ \\
      $|10\,1\,\textrm{-1}\rangle$ & 2 & -2 & Sinking & $0.086$ & $5 \times 10^{-7}$ & $164.1$ & $-0.05$ \\
      $|11\,1\,\textrm{-1}\rangle$ & 2 & -2 & Sinking & $0.063$ & $4.3 \times 10^{-7}$ & $163.3$ & $-0.036$ \\
      $|12\,1\,\textrm{-1}\rangle$ & 2 & -2 & Sinking & $0.047$ & $3.8 \times 10^{-7}$ & $162.8$ & $-0.026$ \\
      $|13\,1\,\textrm{-1}\rangle$ & 2 & -2 & Sinking & $0.036$ & $3.3 \times 10^{-7}$ & $162.4$ & $-0.019$ \\
      $|14\,1\,\textrm{-1}\rangle$ & 2 & -2 & Sinking & $0.028$ & $3 \times 10^{-7}$ & $162.0$ & $-0.014$ \\
      $|3\,0\,\textrm{0}\rangle$ & 1 & -1 & Sinking & $100$ & $1.4 \times 10^{-5}$ & $157.6$ & $-95$ \\
      $|4\,0\,\textrm{0}\rangle$ & 1 & -1 & Sinking & $17$ & $1.1 \times 10^{-5}$ & $123.8$ & $-45$ \\
      $|5\,0\,\textrm{0}\rangle$ & 1 & -1 & Sinking & $5.8$ & $7.9 \times 10^{-6}$ & $113.9$ & $-23$ \\
      $|6\,0\,\textrm{0}\rangle$ & 1 & -1 & Sinking & $2.7$ & $6.1 \times 10^{-6}$ & $109.2$ & $-13$ \\
      $|7\,0\,\textrm{0}\rangle$ & 1 & -1 & Sinking & $1.5$ & $4.8 \times 10^{-6}$ & $106.8$ & $-7.6$ \\
      $|8\,0\,\textrm{0}\rangle$ & 1 & -1 & Sinking & $0.94$ & $4 \times 10^{-6}$ & $105.2$ & $-5$ \\
      $|9\,0\,\textrm{0}\rangle$ & 1 & -1 & Sinking & $0.62$ & $3.3 \times 10^{-6}$ & $104.2$ & $-3.4$ \\
      $|10\,0\,\textrm{0}\rangle$ & 1 & -1 & Sinking & $0.44$ & $2.8 \times 10^{-6}$ & $103.5$ & $-2.4$ \\
      $|11\,0\,\textrm{0}\rangle$ & 1 & -1 & Sinking & $0.32$ & $2.5 \times 10^{-6}$ & $103.0$ & $-1.8$ \\
      $|12\,0\,\textrm{0}\rangle$ & 1 & -1 & Sinking & $0.24$ & $2.2 \times 10^{-6}$ & $102.6$ & $-1.4$ \\
      $|13\,0\,\textrm{0}\rangle$ & 1 & -1 & Sinking & $0.19$ & $1.9 \times 10^{-6}$ & $102.3$ & $-1.1$ \\
      $|14\,0\,\textrm{0}\rangle$ & 1 & -1 & Sinking & $0.15$ & $1.7 \times 10^{-6}$ & $102.1$ & $-0.86$ \\
      \hline
    \end{tabular}
    \caption{The series of Bohr resonances encountered by a $|2\,1\,1\rangle$ cloud with $M_{c,0} = 0.1M$ and $\alpha =0.2$, due to the perturbations from a counter-rotating binary system with $q = 0.1$. We report for each resonantly excited state the multipole terms ($\ell_*,\,g$) involved in the perturbation, the sinking/floating nature of the resonance, the Landau-Zener parameter $Z$, the mixing term at resonance $\eta^{(g)}$ (reported here in dimensionless units, \emph{i.e.} computed by considering $G=c=M=1$), the radius at which the transition is excited $R_\textrm{res}$ (by which we sort the table in descending order) and the fraction $\Delta M_c$ of cloud mass that falls back to the primary black hole due to the $\Gamma$ decay (here we also take into account the backreaction on the orbit and the $\Gamma$ decay of the excited state) when only considering that particular resonance. Here we only show resonances that transfer a fraction of mass above 0.001\% to the primary black hole. The strongest sinking resonances are those to the $n_b=3$ states for both the dipole $(\ell_*=1)$ and the quadrupole ($\ell_*=2$) couplings. Although, for the former case, resonances to states with $n_b > 3$ can significantly alter the cloud mass.}
    \label{tab:resonances}
  \end{table*}

  \begin{figure*}[!t]
    \includegraphics[width=\textwidth]{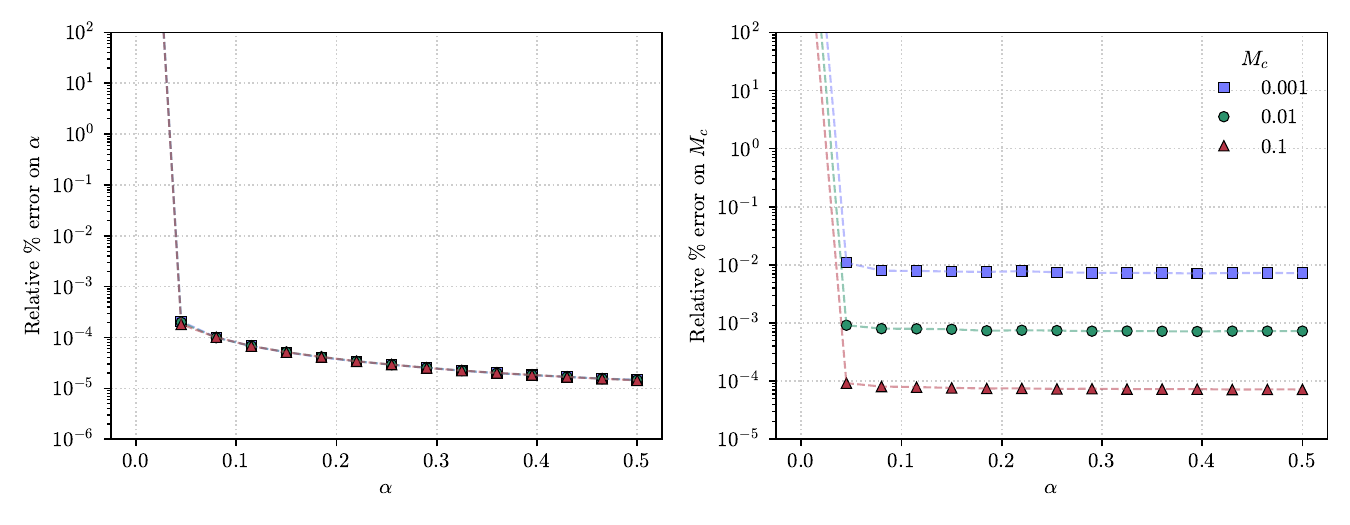}
    \caption{Relative percentage errors on the scalar cloud parameters, $\alpha$ (left panel) and the cloud mass $M_c$ (right panel), for the values of $\alpha$ and of the cloud mass, $M_c$, already considered in Fig.~\ref{fig:ET_fisher_matrix_detectability}, but switching off the resonances in our model. The only appreciable difference with respect to the full-model case is the disappearance of bumps, leading to significantly smoother detectability curves. This suggests that the loss of precision in the full analysis is directly associated with the inclusion of resonant interactions. The effect is most visible for the $M_c = 0.1M$ curve.}
    \label{fig:ET_fisher_matrix_detectability_no_resonances}
  \end{figure*}

  Here we report the details of the resonant Bohr transitions from an initial $|2\,1\,1\rangle$ cloud state, to states with principal quantum numbers $n_b \geq 3$ (up to $n_b = 14$). In particular, focusing on the
  counter-rotating
  binary with a mass ratio of $q = 0.1$, that we have taken as benchmark system throughout this work, we consider resonances driven by both dipole $(\ell_* = 1)$ and quadrupole $(\ell_* = 2)$ couplings.
  We have checked that for resonances mediated only by higher multipoles, $\ell_* > 2$, both the dissipated fraction of cloud mass and the orbital backreaction are strongly suppressed. Nonetheless, for those resonances that can be excited by different multipole modes (e.g. transitions to $|n_b20\rangle$ states that, according to the selection rules in Eqs.~\eqref{eq:selection_1}-\eqref{eq:selection_3}, can have $\ell_*=1,\,3$) we compute the level mixing in Eq.~\eqref{eq:bound-bound-mixing}, by summing over all compatible multipoles.
  The intensity of all the considered resonances is described in Table \ref{tab:resonances}, for an initial cloud mass $M_{c,0} = 0.1M$ and for $\alpha = 0.2$, under the influence of a counter-rotating binary system. For each resonant state excited by the binary's perturbation, we list the contributing multipole terms $(\ell_*, m_*)$, the Landau-Zener parameter $Z$ , the resonance-specific mixing term $\eta^{(g)}$, the resonance radius $R_\textrm{res}$ (sorted in descending order), and the fractional mass $\Delta M_c$ that falls back onto the primary black hole due to the decay rate $\Gamma$ of the excited state. The table is filtered to include only resonances where the mass transfer exceeds 0.001\%, focusing attention on those transitions with the most substantial impact on the cloud's evolution.
  The strongest resonances among those considered here are the transitions to the $|300\rangle$ state and in general all those to $|n_b00\rangle$ states. These resonances, when considered alone, can largely deplete or completely disrupt the cloud.
  However, it is important to remark that the percentages of cloud dissipation shown in the table do not correspond to those that will be produced in the final case when the full dynamics is taken into account. In fact, in these calculations, the main force that drives the inspiral is the emission of gravitational waves, while, as we will show later at short orbital separations corresponding to the cloud size, where the Bohr resonances live, the main driving force of the inspiral is the dynamical friction produced by cloud ionization, which reduces the time the binary spends in the resonance band, thus reducing the dissipative effect. This also reflects in the fact that, when all effects are taken into account and for realistic combinations of cloud parameters, the mass and spin of the central black hole only evolve by a tiny fraction (variations are of order $\lesssim 0.1\%$) and we can thus neglect them.

  \section{Detectability of cloud without resonances}
  \label{app:detectabilty-no-resonances}

  For completeness, we show in Fig.~\ref{fig:ET_fisher_matrix_detectability_no_resonances} the result of our detectability analysis of the boson cloud with ET, obtained by switching off the resonances in our model. The general trends observed in the full analysis are preserved: the parameter estimation improves for increasing values of $\alpha$ and higher cloud masses. However, in contrast to the full analysis, the curves here are much smoother, and the sharp peaks appearing at specific values of $\alpha$, particularly in the $M_c = 0.1M$ case, are entirely absent. This supports the interpretation that the ``bumps'' observed earlier originate from resonant transitions entering or exiting the detector band.

  \end{document}